\documentclass[secnumarabic,amssymb, nobibnotes, aps, prd]{revtex4-2}
\usepackage{graphicx}
\usepackage{amsmath}
\usepackage{subcaption}
\usepackage{microtype}
\usepackage{xcolor}

\usepackage{multirow}

\begin{document}
	
	\title{ BTZ Black Hole In The Non-Extensive Generalizations of Gibbs Entropy }%
	
	\author{Amijit Bhattacharjee$^1$}
	
	\email{$rs_amijitbhattacharjee@dibru.ac.in$}
	
	\author{Prabwal Phukon$^{1,2}$}
	\email{prabwal@dibru.ac.in}
	
	\affiliation{$1.$Department of Physics, Dibrugarh University, Dibrugarh, Assam,786004.\\$2.$Theoretical Physics Division, Centre for Atmospheric Studies, Dibrugarh University, Dibrugarh, Assam,786004.}
	\maketitle
	\section*{Abstract}
	 We study the thermodynamics and thermodynamic geometry of the (2+1) dimensional Banados-Teitelboim-Zanelli(BTZ) black hole within the framework of the non-extensive generalizations of Gibbs entropy. We investigate both the rotating (R-BTZ) and the charged (C-BTZ) BTZ black holes in these non-extensive entropy formalisms. We write down the Bekenstein-Hawking(BH) entropy of the black hole in terms of the non-extensive entropies namely: Kaniadakis entropy, Renyi entropy and Barrow entropy. We investigate their impact on the thermodynamic phase structure and geometry of the BTZ black holes in both the ensembles i.e. the fixed $(J)$ and fixed $(\Omega)$ ensemble for the R-BTZ black hole  and  the fixed $(Q)$  and the fixed $(\Phi)$ ensemble for the C-BTZ black hole  where $J$, $\Omega$, $Q$ and $\Phi$  represent the angular momentum, angular velocity, charge and the electric potential of the respective black holes. We investigate the Ruppeiner and geometrothermodynamic(GTD) geometries of the black hole for all the non-extensive entropy cases. We find that there are Davies type along with Hawking-Page phase transitions in both the charged and rotating  BTZ black hole for the Kaniadakis entropy case in all the above mentioned thermodynamic ensembles. These phase transitions were not seen in the BH entropy case. We also find that the Ruppeiner and the GTD scalar for the Kaniadakis entropy show curvature singularities corresponding to the Davies type phase transitions in both the rotating and charged BTZ black holes.
\section{Introduction:}
It is by now well known that black holes are thermodynamic objects characterized by a temperature (Hawking temperature) proportional to surface gravity at its horizon and entropy (Bekenstein-Hawking entropy) proportional to the horizon area  \cite{hawking1975particle,bekenstein1973black}. 

\begin{equation}
T=\frac{\kappa}{2\pi},~~~~~~~~S_{BH}=\frac{A}{4}
\label{eqtn1}
\end{equation}
These two quantities ($T$ and $S_{BH}$) constitute the basis of what is now known as black hole thermodynamics \cite{hawking1976black}. Over the years, black hole thermodynamics has evolved into an active area of research with a number of generalizations of its originial version \cite{cai2013pv,kubizvnak2017black,visser2022holographic,
cong2021thermodynamics,gao2022restricted,gao2022thermodynamics} shedding light on the phase structures of black holes. \\

An important issue in the context of black hole thermodynamics  that has recently come to focus is the modification of black hole entropy owing to different corrections. These corrections result in a number of  non-extensive generalizations of Gibbs entropy. Some of these notable entropies are:  Kaniadakis entropy,  \cite{kaniadakis2002statistical,drepanou2022kaniadakis}, Renyi entropy \cite{renyi1959dimension} and Barrow entropy \cite{barrow2020area}. A brief introduction to these entropies will be provided in the next section. The impact of these entropies on the thermodynamic behaviour of a few black holes have been studied in a number of works \cite{luciano2023p,promsiri2022emergent,ghaffari2019black,
luciano2023black,jawad2022thermodynamic}. Previously these entropy systems have been tested in cosmology \cite{saridakis2020barrow,saridakis2018holographic, tavayef2018tsallis,nojiri2022nonextensive,nojiri2022early} and quantum physics \cite{nojiri2021area,nojiri2022alternative,shababi2020non,
luciano2021tsallis,moussa2021schwarzschild} with notable success.\\

Although we employ Kaniadakis, Renyi, and Barrow entropies individually for our convenience here, it is worth noting that  generalized entropy constructs with different parameters  have been introduced in \cite{nojiri2022nonextensive,nojiri2022early} that generalizes all the above mentioned entropies. These generalized entropy constructs are formulated such that they reduce to all the required entropy paradigms under appropriate  limits. For example, the six-parameter entropy formulation proposed in \cite{nojiri2022early} is given as:

\begin{equation}
S_G(\alpha_{\pm}, \beta_{\pm}, \gamma_{\pm}) = \frac{1}{\alpha_{+} + \alpha_{-}} \left[ \left( 1 + \frac{\alpha_{+}}{\beta_{+}} S^{\gamma_{+}} \right)^{\beta_{+}} - \left( 1 + \frac{\alpha_{-}}{\beta_{-}} S^{\gamma_{-}} \right)^{-\beta_{-}} \right],
\end{equation}
where all parameters $(\alpha_{\pm}, \beta_{\pm}, \gamma_{\pm})$ are positive. This expression reduces to the above mentioned entropy formulations under specific parameter limits, as has been outlined below:

\begin{itemize}

     \item \textbf{Kaniadakis Entropy:} By taking $\beta_{\pm} \to 0$, $\gamma_{\pm} = 1$, and $\alpha_{\pm} = K$, the generalized entropy simplifies to
    \begin{equation}
    S_G \to \frac{1}{K} \sinh(KS),
    \end{equation}
    which corresponds to the Kaniadakis entropy.

  \item \textbf{Renyi Entropy:} From the Sharma-Mittal limit, if $\alpha_{+} \to 0$, $\beta_{+} \to 0$, and $\alpha_{+}/\beta_{+} \to \alpha$, with $\gamma_{+} = 1$, we recover the Renyi entropy:
    \begin{equation}
    S_G \to \frac{1}{\alpha} \ln (1 + \alpha S).
    \end{equation}
    
  \item \textbf{Sharma-Mittal Entropy:} Setting $\alpha_{-} = 0$ and $\gamma_{+} = 1$, we obtain
    \begin{equation}
    S_G = \frac{1}{\alpha_{+}} \left[ \left( 1 + \frac{\alpha_{+}}{\beta_{+}} S^{\gamma_{+}} \right)^{\beta_{+}} - 1 \right].
    \end{equation}
    Choosing $\alpha_{+} = R$, $\beta_{+} = R/\delta$, and $\gamma_{+} = \delta$, the Sharma-Mittal entropy is reproduced.
    
     \item \textbf{Tsallis and Barrow Entropy:} In the limit $\alpha_{+} = \alpha_{-} \to 0$ and $\gamma_{+} = \gamma_{-} = \gamma$, the generalized entropy reduces to $S_G \to S^{\gamma}$. By further choosing $\gamma = \delta$ or $\gamma = 1 + \Delta/2$, we recover the Tsallis entropy and Barrow entropy, respectively.
    
   \item \textbf{Loop Quantum Gravity Entropy:} Setting $\alpha_{-} = 0$ and $\gamma_{+} = 1$, the entropy becomes
    \begin{equation}
    S_G = \frac{1}{\alpha_{+}} \left[ e^{\beta_{+} S} - 1 \right].
    \end{equation}
    In the limit $\beta_{+} \to \infty$ and $\alpha_{+} = 1 - q$, this reduces to the Loop Quantum Gravity entropy
    \begin{equation}
    S_G \approx \frac{1}{1 - q} \left[ e^{(1-q)S} - 1 \right],
    \end{equation}
    which approaches the Bekenstein-Hawking entropy as $q \to 1$.
\end{itemize}

The entropy $S_G$ defined above do satisfy the generalized third law of thermodynamics and is seen to vanish when $S \to 0$. $S_G$ is found to be a monotonically increasing function of $S$, thus reducing to the Bekenstein-Hawking entropy under suitable limits.\\

Although black hole thermodynamics in its different forms have largely been successful in providing deep insights about the thermodynamic properties of black holes, there are a number of important issues which are far from being  completely understood. One such issue is the microscopic origin of Bekenstein-Hawking(BH) entropy. In absence of  a complete theory of quantum gravity, thermodynamic geometry studies of black holes have been used to extract qualitative ideas about microscopic interaction in black holes \cite{wang2020thermodynamic,shahzad2022study,
wu2021ruppeiner}. Thermodynamic geometry is a key concept in the study of black holes as they provide an understanding of the phase transitions in terms of the curvature singularities obtained through a particular thermodynamic metric. Three such metrics which have been used in black hole thermodynamics are Weinhold, Ruppeiner and geometrothermodynamic metric (GTD).\\

Weinhold \cite{weinhold1975metric} introduced a Riemannian metric in the equilibrium space which is defined as the second derivative of the internal energy with respect to other extensive variables. The Weinhold metric is defined as:
$$g_{ij}^{W} = \partial_{i} \partial_{j} U(S,X)$$  
where the internal energy `$U(S,X)$' is a function of entropy S and other extensive variables X. \\

Ruppeiner \cite{ruppeiner1979thermodynamics} later introduced a Riemannian metric which is defined as the negative hessian of the entropy with respect to the other extensive variables. The Ruppeiner metric is a concept based on the thermodynamic fluctuation theory of equilibrium thermodynamics, where the thermodynamic length is related to the probability of fluctuation between two states of the thermodynamic system \cite{salamon1984relation}. The two are inversely related as an increase in the probability of fluctuation  would mean a significant decrease in the thermodynamic length and vice-versa. The Ruppeiner metric is defined as:
 $$g_{ij}^{R} =- \partial_{i} \partial_{j} S(U,X)$$
where the entropy `$S(U,X)$' is a function of internal energy $U$ and other extensive variables X. The Riemannian scalar which is defined here as `$R_{rupp}$' is a scalar invariant function in thermodynamic geometry. The sign of the scalar curvature has been linked to the nature of microscopic interactions in the thermodynamic system. For positive or negative scalar curvatures, the underlying interaction is either repulsive or attractive accordingly. Whereas the null curvature would mean an absence of interaction with a flat thermodynamic geometry. The Ruppeiner and the Weinhold metrics are conformally related  to each other by the following relation \cite{ruppeiner1995riemannian,mrugala1984equivalence}:
\begin{equation}
g_{ij}^{R} dU^i dU^j = \frac{1}{T}  g_{ij}^{W} dS^i dS^j
\label{eqtn2}
\end{equation}
where $T=\frac{\partial{U}}{\partial{S}}$ is the Hawking temperature of the system with $U^i =(U,X)$ and $S^i =(S,X)$.\\

The Ruppeiner and Weinhold geometries have successfully predicted the occurrence of phase transitions in the case of ordinary thermodynamic systems. However there has been contradictory results for the case of black holes where for instance in Kerr-black hole \cite{aaman2003geometry} the Weinhold metric showed no phase transition whereas the Ruppeiner metric could show phase transitions only for a specific thermodynamic potential. However such problems were successfully addressed in the new geometric formalism of geometrothermodynamics(GTD)\cite{quevedo2007geometrothermodynamics} where the properties of the phase space and the space of equilibrium states could be unified \cite{quevedo2008geometrothermodynamics}. The GTD metric is a Legendre invariant metric and therefore would not depend on any specific choice of thermodynamic potential. The phase transitions obtained from the specific heat capacity of the black hole are properly contained in the scalar curvature of the GTD metric, such that a curvature singularity in the GTD scalar `$R_{GTD}$'  would imply the occurrence of a phase transition. The general form of the GTD metric is given by \cite{soroushfar2016thermodynamic}:
\begin{equation}
g = \left(E^{c} \frac{\partial{\varphi}}{\partial{E^{c}}} \right)
\left(\eta_{ab} \delta^{bc} \frac{\partial^2 \varphi}{\partial E^{c} \partial E^{d}} dE^{a} dE^{d}\right)
\label{eqtn3}
\end{equation}     
where `$\varphi$' is the thermodynamic potential and `$E^a$' is an extensive thermodynamic variable with $a=1,2,3....$. \\

In this paper, we study the thermodynamics and thermodynamic geometry of the (2+1) dimensional Banados-Teitelboim-Zanelli(BTZ) black hole within the framework of the non-extensive generalizations of Gibbs entropy. We investigate both the rotating (R-BTZ) and the charged (C-BTZ) BTZ black holes in these non-extensive entropy formalisms. We write down the Bekenstein-Hawking(BH) entropy of the black hole in terms of the non-extensive entropies namely: Kaniadakis entropy, Renyi entropy and Barrow entropy. We investigate their impact on the thermodynamic phase structure and geometry of the BTZ black hole in both the ensembles i.e. the fixed $(J)$ and fixed $(\Omega)$ ensemble for the R-BTZ black hole  and  the fixed $(Q)$  and the fixed $(\Phi)$ ensemble for the C-BTZ black hole  where $ J$, $\Omega$, $Q$ and $\Phi$  represent the angular momentum, angular velocity, charge and the electric potential of the respective black holes. We investigate the Ruppeiner and geometrothermodynamic(GTD) geometries of the black hole for all the non-extensive entropy cases.\\

In this study, we adopt the energy definition of General Relativity, the first law of thermodynamics and the conventional definition of thermodynamic temperature as fundamental. However we accept the alternative viewpoint that has
been adopted in \cite{nojiri2021area,nojiri2022alternative} based on the assumption that the Hawking temperature must be kept unaffected. Furthermore, it has been demonstrated  that non-extensive entropy paradigms can lead to wrong expressions for the Hawking temperature or black hole energy that differ from the standard ones.  While we acknowledge these critiques, we argue that  any modification to these quantities such as the Hawking temperature under these non-extensive entropy formalisms, should be interpreted as a shift in the effective thermodynamic description rather than a fundamental change
to the Hawking temperature itself. These effective temperatures reflect the corrections introduced by non-extensive statistical mechanics and are not meant to replace the fundamental Hawking temperature. The effective temperatures characterize the modified thermodynamic equilibrium within each of the non-extensive paradigms and provide insights into how such generalizations affect black hole systems. These differences highlight the speculative yet potentially valuable role of non-extensive
frameworks in exploring new physical scenarios and enriching our understanding of black hole thermodynamics. We stress upon the fact that the exploration of different non-extensive entropies is motivated by their success in describing physical systems beyond black holes
\cite{silva2006relativistic,yarahmadi2024using,mcminis2013renyi,tu2022renyi,
leon2021barrow,feng2023barrow}. The study of their implications for black hole thermodynamics provides us with an opportunity to test the limits of our current understanding and to further explore potential extensions of the standard approach.\\

The paper is structured as follows: In section \textbf{II} we review some known facts about the non-extensive generalisations of  Gibbs entropy. The section \textbf{III} is about the rotating BTZ black hole and its thermodynamic investigation within the framework of non-extensive entropies in two different ensembles. In section \textbf{IV} we examine the thermodynamic geometry of the black hole system both in Ruppeiner and in the GTD formalisms. The section \textbf{V} is about the thermodynamic investigation of charged BTZ black hole  within the framework of non-extensive entropies in two different ensembles. In section \textbf{VI} we examine the thermodynamic geometry of the black hole in both the Ruppeiner and  the GTD formalisms. The final section contains our conclusions.

\section{Non-extensive generalizations of gibbs entropy}

The entropy of a black hole scales with its area rather than its volume and is therefore regarded as a non-extensive quantity. Due to its non-extensive nature, the black hole entropy is also non-additive and follows a non-additive composition rule. But in the case of Gibbs thermodynamics the entropy is defined to be both extensive and additive as it scales with the size of the system. However, the assumption of extensivity of Gibbs entropy is due to ignoring the long range forces that are prevalent between the thermodynamic sub-systems. Gibbs thermodynamics ignores these forces because the size of the system exceeds the interaction range between the thermodynamic sub-systems. The total entropy thus becomes equal to the sum of the entropies of its components. It therefore grows with the size of the thermodynamic system.\\

But these long range forces are important in various thermodynamic systems such as black holes. We can therefore infer that Gibbs thermodynamics may not be a suitable choice for studying black hole thermodynamics. Therefore in order to understand the non-extensive nature of black hole entropy, several non-extensive generalizations of Gibbs entropy have been prescribed. We give a brief outline of a few of those entropies below which we will use later in order to study the thermodynamics and thermodynamic geometry of both the rotating  and charged BTZ black holes respectively:  

\subsubsection*{\textbf{A. Kaniadakis entropy}}
 Kaniadakis entropy\cite{kaniadakis2002statistical,drepanou2022kaniadakis,
luciano2023p, kaniadakis1996generalized,
 kaniadakis2004deformed,
 kaniadakis2005statistical,housset2024cosmological} is a relativistic non-extensive generalization of Boltzmann-Gibbs entropy. It was proposed by Kaniadakis so as to accomodate special relativity with non-extensive statistical mechanics. Kaniadakis entropy is premised upon the fact that physical observables such as momentum and energy when generalised under special relativity are viewed as a one-parameter deformation of the corresponding non-relativistic formula i.e $P' =\frac{P}{\sqrt{1- x^2}}$ where, $P'$ is the relativistic momentum and $x=\frac{v}{c}$ is the deformation parameter. It is therefore quite obvious that the entropy of a relativistic system could also be seen as a one parameter deformation of the classical non-relativistic entropy. In order to achieve a relativistic formula for Boltzmann Gibbs entropy, Kaniadakis proposed a deformed logarithmic function in the entropy formula which in turn deforms the Maxwell-Boltzmann distribution function. These deformed functions are given as:
 
  $$\ln_K (x) = \frac{x^K - x^{-K}}{2 K} $$
 
 $$ exp_K (x)= \left(\sqrt{1 + K^2 x^2} + K x\right)^\frac{1}{K}$$

 where x is a variable and K is the deformation parameter called the Kaniadakis parameter and both the above mentioned deformed functions reduce to their original form as the deformed parameter, K approaches appropriate limit i.e. $K \rightarrow 0$. The Kaniadakis entropy is then defined as:
 $$S_{K}=-\sum_{i=1} ^ {n} P_{i} \ln_K (P_{i})$$
 where,
  $$\ln_K (P_{i}) = \frac{P_{i}^K - P_{i}^{-K}}{2 K} $$
  In the limit $K \rightarrow 0$, the Kaniadakis entropy reduces to the Boltzmann-Gibbs entropy. Considering the equiprobability of microstates inside the black hole we can reduce the Kaniadakis entropy to:
  $$S_{K}=-\sum_{i=1} ^ {n} \frac{1}{n} \ln_K \left(\frac{1}{n}\right)$$ 
  $$S_{K} = \ln_K (n)$$
 where `$n$' is the number of microstates of the thermodynamic system and `K' is the kaniadakis parameter. On further simplification we obtain:
  $$S_{K} = \frac{n^K - n^{-K}}{2 K}$$
  $$S_{K} = \frac{Sinh[K \ln (n)]}{ K}$$
 Considering that the Bekenstein-Hawking entropy is obtained by counting the number of microstates inside the black hole as was showed by Strominger and Vafa in their seminal work \cite{strominger1996microscopic}, we write the Bekenstein -Hawking entropy, $S_{BH}=\ln (n) $ such that the above equation reduces to:
 \begin{equation}
 S_{K}=\frac{Sinh[K S_{BH}]}{K}
 \label{eqtn4}
 \end{equation}
 where, for the limit $K \rightarrow 0$, $S_{K} \rightarrow S_{BH}$
 \subsubsection*{\textbf{B. Renyi entropy}}
 The Renyi entropy\cite{nakarachinda2022thermodynamics,lenzi2000statistical,
 jizba2004world,jizba2004observability,dowker2013sphere,
 liu2013refinement,fursaev2012entanglement} is a non-extensive generalization of Boltzmann-Gibbs entropy where unlike the case where all the probabilities are treated uniformly, Renyi entropy allows for a parameter adjustment to the sensitivities of different probabilities in the system. This becomes very useful in situation where either rare or high probable events demand separate emphasis. The Renyi entropy is a measure of the quantum entanglement of a system. A phenomena where the quantum states of two or more particles could become correlated. 
 It is defined as:
 $$S_{R}=\frac{1}{1-q} \left(\ln\sum_{i=1} ^ {n} P^{q}(i)\right)$$
 where `P(i)' is the probability distribution and `q' is the non-extensive Tsallis parameter. We here assume that the black hole entropy is just the Tsallis entropy \cite{tsallis2013black,furuichi2006information,
 dos1997generalization} which is a one parameter generalization of Boltzmann-Gibbs entropy so as to include both extensive and non-extensive statistical systems. The Tsallis entropy is given by :
 $$ S_T = S_{BH} = \frac{1}{1-q} \left(\sum_{i=1} ^ {n} P^{q}(i) - 1\right)$$
 where for $q \rightarrow 1$, $S_T$ reduces to the usual Boltzmann-Gibbs entropy. The parameter \( q \) is responsible for the degree of non-extensivity of the entropy:

\begin{itemize}
    \item \textbf{For \( q = 1 \)}: Tsallis entropy reduces to the  extensive Boltzmann-Gibbs entropy which is appropriate for the system that comprises of weak correlations and short-range interactions.
    \item \textbf{For \( q < 1 \)}: The entropy is sub-extensive and is useful in describing systems where the probability for the occurrence of rare states is higher than the occurence of frequent states.
    \item \textbf{For \( q > 1 \)}: The entropy shows super-extensive behaviour where frequent states do contribute more to the entropy. 
\end{itemize}
By performing  necessary subsitutions, the Renyi entropy could be defined as follows:
 $$ S_R =\frac{\ln \left[ (1-q)S_{BH} +1\right]}{1-q}  $$
 and after putting $1-q =\lambda$ we get the Renyi entropy in terms of Bekenstein-Hawking entropy as:
 \begin{equation}
 S_{R}=\frac{\ln[1 + \lambda S_{BH}]}{\lambda}
 \label{eqtn5}
 \end{equation}
 where `$\lambda$' is the Renyi parameter and for the limit $\lambda \rightarrow 0$, $S_{R} \rightarrow S_{BH}$
 
\subsubsection*{\textbf{C. Barrow entropy}}
The Barrow entropy\cite{barrow2020area,rani2023impact,barrow1982chaotic}, as introduced by John Barrow is proposed to measure the entropy of a black hole system whose smooth horizon has been replaced by the rough fractal structures. Unlike the above two entropy formalisms, Barrow entropy does not have a statistical root. It is a modification introduced in the context of quantum gravity and the fractal structure of space-time. It is formulated by considering quantum-gravitational effects that modify the surface of the black hole's event horizon by considering the structure of the horizon to be fractal-like. John Barrow's approach to entropy suggests that the horizon area might have a fractal dimension which alters the traditional Bekenstein-Hawking entropy formula. This modification puts forth a parameter that can quantify the degree of deviation from the standard horizon geometry due to quantum corrections while capturing the complexity of space-time at smaller scales. The Barrow entropy is formulated as:
\begin{equation}
S_{\Delta}= (S_{BH})^{1 + \Delta/2}
\label{eqtn8}
\end{equation}
where, `$\Delta$' is the Barrow parameter that is linked to the  fractal structure of the system with range $0<\Delta\leq2$. where $\Delta\rightarrow2$ would yield maximal fractal structure. And for the limit $\Delta \rightarrow 0$,  $S_{\Delta} \rightarrow S_{BH}$ 
\section{Thermodynamics of the  rotating BTZ black hole}
  The action which facilitates the field equations\cite{banados1992black,clement1993classical,
clement1996spinning,kim1999spinning,hennigar2020rotating,
maeda2024charged}
from which the $(2+1)$ dimensional rotating BTZ black hole solutions are obtained is given as:
$$I=\frac{1}{2\pi} \int d^3 x \sqrt{-g}\biggl(R - 2\Lambda\biggr) $$
where, the AdS length $l$ is related to the cosmological constant $\Lambda$ by the relation: $\Lambda=- \frac{1}{l^2}$ and the Einstein's field equations are given by:
$$G_{\mu \nu} - \Lambda g_{\mu \nu}= 0$$

The line element that corresponds to the given solution is:
$$ds^2 = -f(r) dt^2 + \frac{dr^2}{f(r)} + r^2 \biggl(d\phi-\frac{J}{2 r^2} dt\biggr)^2$$
where $f(r) = -M + \frac{r^2}{l^2} + \frac{J^2}{4 r^2}$ and `M' and `J'  are the mass and angular momentum  carried by the black hole. Solving the function for $f(r)=0$ would give us the roots which determine the horizon radius. For an exterior horizon radius, `$r_{+}$' the black hole mass is given by:
\begin{equation}
M=\frac{r_{+}^2}{l^2} + \frac{J^2}{4 r_{+}^2}
\label{eqtn9}
\end{equation}
and the black hole entropy is given by:
$$S_{BH}= 4 \pi r_{+}$$
\subsection{The Kaniadakis entropy case}
The Kaniadakis entropy, $S_K$ in terms of the black hole entropy, $S_{BH}$ is given by (\ref{eqtn4}) as:
$$S_{K}= \frac{Sinh(K S_{BH})}{K} $$
where `K' is the Kaniadakis parameter and for $K \rightarrow 0$, $S_K \rightarrow S_{BH}$. Starting with the above equation we get the modified horizon radius for the CR-BTZ black hole as:
$$S_{K}=\frac{Sinh(4 \pi K r_{+})}{K} $$

 \begin{equation}
 r_{+} = \frac{Arc Sinh(K S_{K})}{4 \pi K}
 \label{eqtn10}
 \end{equation}\\
 
 \underline{\textbf{Fixed $(J)$ ensemble}}: \\
 
 We replace the horizon radius in  (\ref{eqtn9}) by  the one obtained in (\ref{eqtn10}) to get the modified mass $M_{K}$ for the R-BTZ black hole which is given by :
$$M_{K}= \frac{4 \pi^2 K^2 J^2}{ArcSinh^2 [K S_{K}]} + \frac{ArcSinh^2 [K S_{K}]}{16 \pi^2 l^2 K^2} $$
 The function, $ArcSinh[K S_{K}]$ can be replaced by its logarithmic form, $\ln \biggl[KS_{K}+ \sqrt{1 + K^2 S_{K} ^2}\biggr]$ and therefore the expression for mass becomes:
 \begin{equation}
M_{K}= \frac{4 \pi^2 K^2 J^2}{\ln [KS_{K}+ \sqrt{1 + K^2 S_{K} ^2}]^2} + \frac{\ln [KS_{K}+ \sqrt{1 + K^2 S_{K} ^2}]^2}{16 \pi^2 l^2 K^2} 
\label{eqtn11}
\end{equation}\\
 
The  heat capacity for the Kaniadakis entropy case in the fixed $(J)$ ensemble can be obtained as:
\begin{equation}
C_{K} =  \frac{\frac{\partial M_{K}}{\partial S_{K}}}{\frac{\partial^2 M_{K}}{\partial S_{K} ^2}}=- \frac{A}{B}
\end{equation}
where,
\begin{equation}
A= \left( \ln \biggl( K S_K + \sqrt{1 + K^2 S_K^2}\biggr) \biggl( 64 J^2 K^4 \pi^4 - \ln \biggl( K S_K + \sqrt{1 + K^2 S_K^2} \biggr)^4 \biggl( 1 + K^2 S_K^2 \biggr) \biggr)\right)
\end{equation}
and
\begin{align}
B = K \biggl( 
    & 64 S_K J^2 K^5 \pi^4 
      \ln \biggl( K S_K + \sqrt{1 + K^2 S_K^2} \biggr) 
    - K S_K 
      \ln \biggl( K S_K + \sqrt{1 + K^2 S_K^2} \biggr)^5 
     + 192 J^2 K^4 \pi^4 \sqrt{1 + K^2 S_K^2} \notag \\
    & + \ln \biggl( K S_K + \sqrt{1 + K^2 S_K^2} \biggr)^4 
      \sqrt{1 + K^2 S_K^2} 
\biggr)
\end{align}

\begin{figure}[h!t]
		\centering
		\includegraphics[width=9cm,height=6cm]{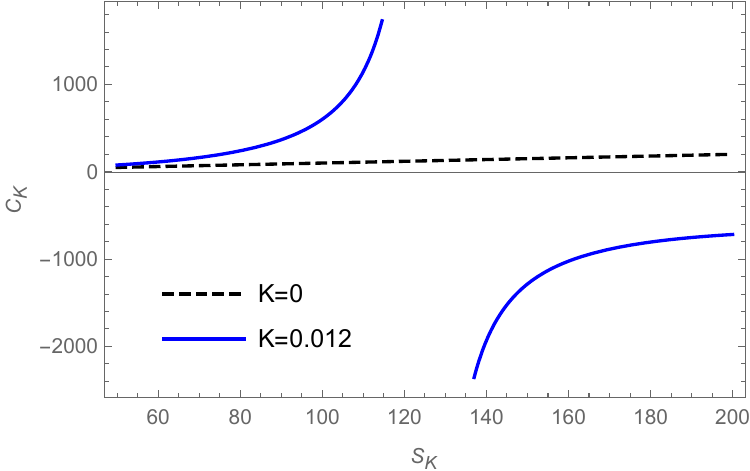}
		\caption{The heat capacity versus Kaniadakis entropy plot in the fixed $(J)$ ensemble for $l=1$ and $J=1$}
		\label{1}
		\end{figure}

In Fig.\ref{1} the heat capacity, $C_K$ is plotted against the Kaniadakis entropy, $S_K$ for $l=1$ and $J=1$ in the fixed $(J)$ ensemble. Here the solid blue curve represents the heat capacity for the Kaniadakis parameter, $K=0.012$ whereas the black dashed line represents the heat capacity for the Bekenstein-Hawking(BH) entropy case $(K=0)$. We find that for $K=0.012$, the heat capacity has a divergence at $S_K=125.77$ which indicates  the presence of a Davies type phase transition present in the Kaniadakis modified black hole whereas the heat capacity for the  BH entropy case shows no such behaviour.	
\\ 
		
		The  Gibbs free energy for the Kaniadakis entropy case in the fixed $(J)$ ensemble can be obtained as:

$$G_{K} =   M_{K} - T_{K} S_{K}$$

\begin{equation}
G_{K} = \frac{\mathcal{A}}{
    16 K^2 \pi^2 \sqrt{1 + K^2 S_{K}^2} \ln\biggl(K S_{K} + \sqrt{1 + K^2 S_{K}^2}\biggr)^3}
\end{equation}\\

where,
\begin{align}
\mathcal{A} = &\ 128 J^2 K^5 \pi^4 S_{K} 
    + 64 J^2 K^4 \pi^4 \sqrt{1 + K^2 S_{K}^2} 
      \ln\biggl(K S_{K} + \sqrt{1 + K^2 S_{K}^2}\biggr)
     - 2 K S_{K} 
      \ln\biggl(K S_{K} + \sqrt{1 + K^2 S_{K}^2}\biggr)^4 \notag \\
    & + \sqrt{1 + K^2 S_{K}^2} 
      \ln\biggl(K S_{K} + \sqrt{1 + K^2 S_{K}^2}\biggr)^5.
\end{align}\\

\begin{figure}[h]	
	\centering
	\begin{subfigure}{0.37\textwidth}
		\includegraphics[width=\linewidth]{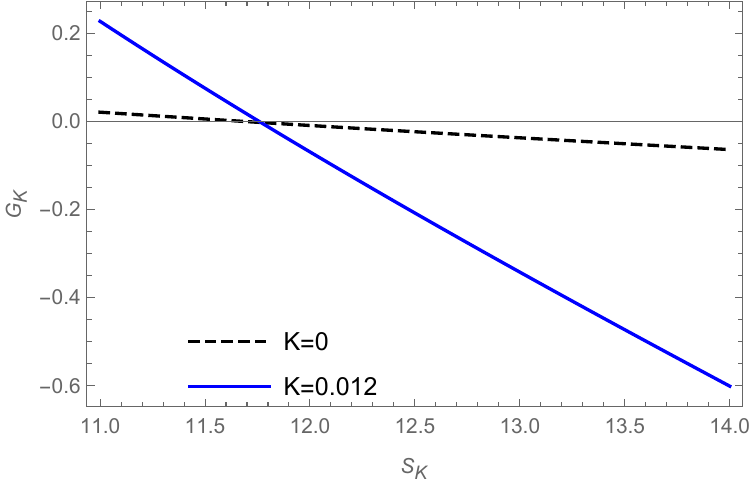}
		\caption{Gibbs free energy versus Kaniadakis entropy $(11< S_K <14)$.}
		\label{2a}
		\end{subfigure}
		\begin{subfigure}{0.37\textwidth}
		\includegraphics[width=\linewidth]{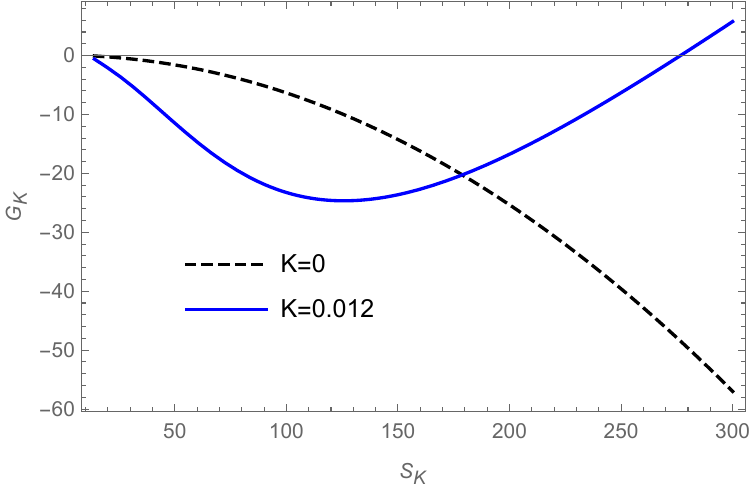}
		\caption{Gibbs free energy versus Kaniadakis entropy $(S_K >14)$.}
		\label{2b}
		\end{subfigure}
		\begin{subfigure}{0.40\textwidth}
		\includegraphics[width=\linewidth]{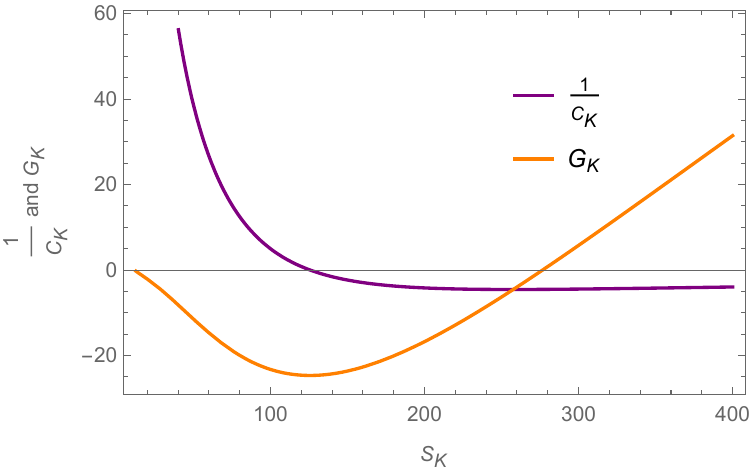}
		\caption{ Inverse specific heat and Gibbs free energy versus Kaniadakis entropy.}
		\label{2c}
		\end{subfigure}
		\caption{The Gibbs free energy curves for Kaniadakis entropy in  the fixed $(J)$ ensemble for $l=1$ and $J=1$.}
	\label{2}
 \end{figure}

In Fig.\ref{2a} and Fig.\ref{2b} the Gibbs free energy, $G_K$ is plotted against the Kaniadakis entropy, $S_K$ for $l=1$ and $J=1$ in the fixed $(J)$ ensemble. We draw two different plots so as to make the presence of the phase transitions more transparent. Here the solid blue curve represents the Gibbs free energy for the Kaniadakis parameter, $K=0.012$ whereas the black dashed line represents the free energy for the Bekenstein-Hawking(BH) entropy case $(K=0)$. The free energy for the $K=0$ case has been scaled down by multiplying it with $\frac{1}{10}$  whereas the free energy for the $K=0.012$ case is kept the same, this difference in scaling is just so that the distinct behaviour of both the curves could be seen more clearly. We keep in mind that the scaling does not affect the position or occurrence of the phase transitions. We find that in Fig.\ref{2a} the Gibbs free energy goes to zero for both the entropy cases, where for $K=0$, the zero point is at $S_K =11.69$ and for $K=0.012$, the zero point is found to be at $S_K =11.75$, these zero points refer to the presence of a Hawking-Page phase transition present in the rotating BTZ black hole for both the Bekenstein-Hawking as well as the Kaniadakis entropy cases. We again find that for $K=0.012$, the Gibbs free energy in Fig.\ref{2b} has a zero point at $S_K=276.64$ only for the Kaniadakis entropy case which indicates  the presence of a Hawking-Page phase transition present in the Kaniadakis modified black hole whereas the Gibbs free energy for the  BH entropy case shows no such behaviour. In  Fig.\ref{2c} we plot the inverse of specific heat capacity (purple) and the Gibbs free energy (orange) against the Kaniadakis entropy. The inverse of specific heat capacity  has been scaled up by multiplying it with $3000$  whereas the free energy is kept the same, this difference in scaling is just so that the distinct behaviour of both the curves could be seen more clearly. We keep in mind that the scaling does not affect the position or occurrence of the phase transitions. The points where these curves intersect the horizontal axis would represent the Davies-type (Purple) and Hawking-Page (Orange) phase transition respectively which are seen to be quite distant from each other.   \\	
 
 \underline{\textbf{Fixed $(\Omega)$ ensemble}}: \\
 
  The modified mass for the Kaniadakis entropy case in the fixed $(\Omega)$ ensemble is given by:
 $$M_{K} ' = M_{K} - J \frac{\partial M_{K}}{\partial J}$$
 where $\frac{\partial M_{K}}{\partial J}= \Omega$ is the angular velocity of the black hole with respect to the angular momentum J. By writing the term $M_{K} '$ as $M_{K}$ for convenience we get:
 \begin{equation}
M_{K} =  \frac{(1 - \Omega^2) \ln\left(K S_{K} + \sqrt{1 + K^2 S_{K}^2}\right)^2}{16 K^2 \pi^2}
\label{eqtn14}
 \end{equation}
  The  heat capacity for the Kaniadakis entropy case in the fixed $(\Omega)$ ensemble can be obtained as:
  $$C_{K} =  \frac{\frac{\partial M_{K}}{\partial S_{K}}}{\frac{\partial^2 M_{K}}{\partial S_{K} ^2}}\\= \frac{\left(1 + K^2 S_K^2\right) \ln\left(K S_K + \sqrt{1 + K^2 S_K^2}\right)}{K \left(\sqrt{1 + K^2 S_K^2} - K S_K \ln\left(K S_K + \sqrt{1 + K^2 S_K^2}\right)\right)}
$$\\

\begin{figure}[h!t]
		\centering
		\includegraphics[width=9cm,height=6cm]{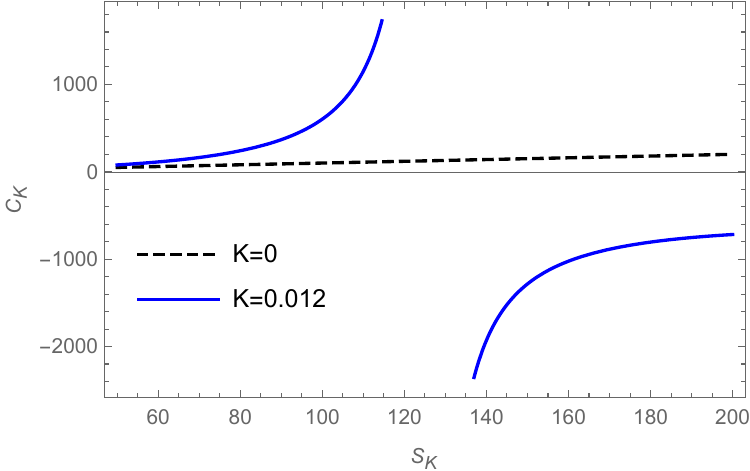}
		\caption{The heat capacity versus Kaniadakis entropy plot in the fixed $(\Omega)$ ensemble for $l=1$ and $\Omega=0.2$}
		\label{3}
		\end{figure}

The heat capacity of the rotating BTZ black hole is independent of $\Omega$ as can be seen from the above expression and therefore the presence or absence of Davies type phase transitions in fixed $(\Omega)$ ensemble is independent of $\Omega$. In Fig.\ref{3} the heat capacity, $C_K$ is plotted against the Kaniadakis entropy, $S_K$ for $l=1$ in the fixed $(\Omega)$ ensemble. Here the solid blue curve represents the heat capacity for the Kaniadakis parameter, $K=0.012$ whereas the black dashed line represents the heat capacity for the Bekenstein-Hawking(BH) entropy case $(K=0)$. We find that for $K=0.012$, the heat capacity has a divergence at $S_K=125.74$ which indicates  the presence of a Davies type phase transition present in the Kaniadakis modified black hole whereas the heat capacity for the  BH entropy case shows no such behaviour. 
  \\
		
		The  Gibbs free energy for the Kaniadakis entropy case in the fixed $(\Omega)$ ensemble can be obtained as:

$$G_{K} =   M_{K} - T_{K} S_{K}$$

\begin{equation}
G_{K} = \frac{(1 - \Omega^2) \ln(K S_{K} + \sqrt{1 + K^2 S_{K}^2}) \left(-2 K S_{K} + \sqrt{1 + K^2 S_{K}^2} \, \ln(K S_{K} + \sqrt{1 + K^2 S_{K}^2})\right)}{16 K^2 \pi^2 \sqrt{1 + K^2 S_{K}^2}}
\end{equation}\\

\begin{figure}[h]	
	\centering
	\begin{subfigure}{0.37\textwidth}
		\includegraphics[width=\linewidth]{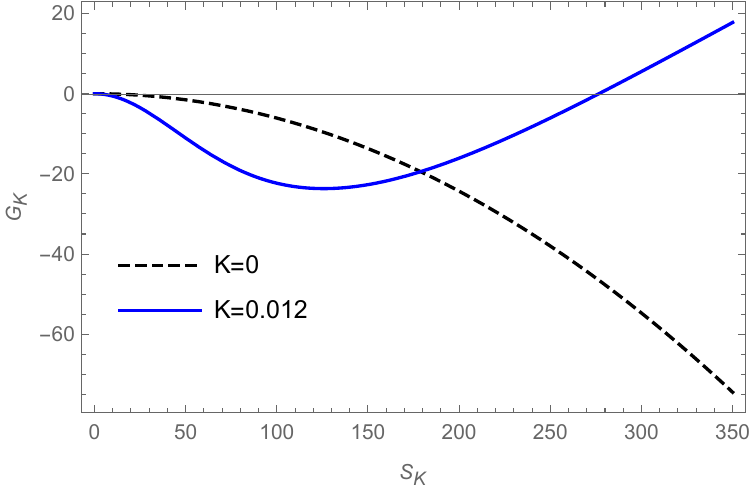}
		\caption{Gibbs free energy versus Kaniadakis entropy.}
		\label{4a}
		\end{subfigure}
		\begin{subfigure}{0.40\textwidth}
		\includegraphics[width=\linewidth]{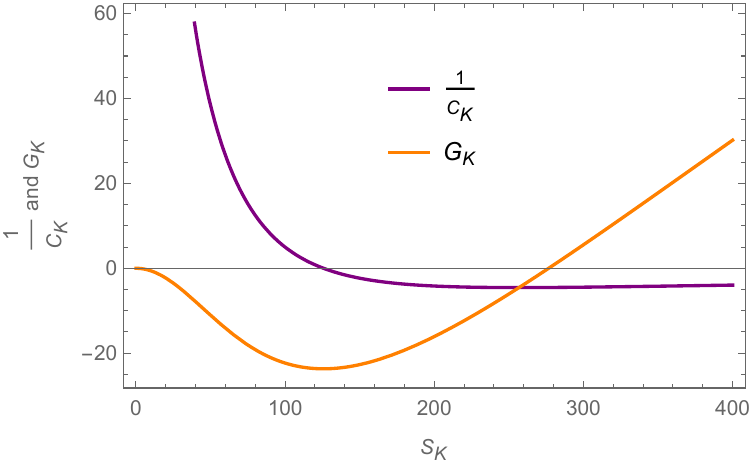}
		\caption{ Inverse specific heat and Gibbs free energy versus Kaniadakis entropy.}
		\label{4c}
		\end{subfigure}
		\caption{The Gibbs free energy curves for Kaniadakis entropy in  the fixed $(\Omega)$ ensemble for $l=1$ and $\Omega=0.2$.}
	\label{4}
 \end{figure}

In Fig.\ref{4a} the Gibbs free energy, $G_K$ is plotted against the Kaniadakis entropy, $S_K$ for $l=1$ and $\Omega=0.2$ in the fixed $(\Omega)$ ensemble. Here the solid blue curve represents the Gibbs free energy for the Kaniadakis parameter, $K=0.012$ whereas the black dashed line represents the free energy for the Bekenstein-Hawking(BH) entropy case $(K=0)$. The free energy for the $K=0$ case has been scaled down by multiplying it with $\frac{1}{10}$  whereas the free energy for the $K=0.012$ case is kept the same, this difference in scaling is just so that the distinct behaviour of both the curves could be seen more clearly. We keep in mind that the scaling does not affect the position or occurrence of the phase transition. We find that for $K=0.012$, the Gibbs free energy has a zero point at $S_K=276.65$ which indicates  the presence of a Hawking-Page phase transition present in the Kaniadakis modified black hole whereas the Gibbs free energy for the  BH entropy case shows no such behaviour. In  Fig.\ref{4c} we plot the inverse of specific heat capacity (purple) and the Gibbs free energy (orange) against the Kaniadakis entropy. The inverse of specific heat capacity  has been scaled up by multiplying it with $3000$  whereas the free energy is kept the same, this difference in scaling is just so that the distinct behaviour of both the curves could be seen more clearly. We keep in mind that the scaling does not affect the position or occurrence of the phase transitions. The points where these curves intersect the horizontal axis would represent the Davies-type (Purple) and Hawking-Page (Orange) phase transition respectively which are seen to be quite distant from each other.   \\	

		\subsection{The Renyi entropy case}
	The Renyi entropy, $S_R$ in terms of the black hole entropy, $S_{BH}$  is given by (\ref{eqtn5}) as:
	 $$S_{R}=\frac{\ln[1 + \lambda S_{BH}]}{\lambda}$$
  where `$\lambda$' is the Renyi parameter and for the limit $\lambda \rightarrow 0$, $S_{R} \rightarrow S_{BH}$. Starting with the above equation we get the modified horizon radius for the R-BTZ black hole as:
  	$$S_{R}=\frac{\ln(1 + 4 \pi \lambda r_{+} )}{\lambda} $$

 \begin{equation}
 r_{+} = \frac{(e^{\lambda S_R} -1)}{4 \pi \lambda}
 \label{eqtn15}
 \end{equation}
 \underline{\textbf{Fixed $(J)$ ensemble}}: \\
  We replace the horizon radius in  (\ref{eqtn9}) by  the one obtained in (\ref{eqtn15}) to get the modified mass $M_{R}$ for the R-BTZ black hole which is given by :
  \begin{equation}
 M_{R} =\frac{4 \pi^2 \lambda^2 J^2}{\left(e^{\lambda S_R}-1\right)^2}+ \frac{\left(e^{\lambda S_R}-1\right)^2}{16 \pi^2 \lambda^2 l^2}
 \end{equation} \\

 The heat capacity for the Renyi entropy case in the fixed $(J)$ ensemble can be calculated as:
$$C_{R}=  \frac{\frac{\partial M_{R}}{\partial S_{R}}}{\frac{\partial^2 M_{R}}{\partial S_{R} ^2}}\\= 
\frac{A}{B}
$$\\
where,
\begin{equation}
A=(-1 + e^{\lambda S_{R}}) \Big(1 - 4 e^{\lambda S_{R}} + 6 e^{2 \lambda S_{R}} - 4 e^{3 \lambda S_{R}} + e^{4 \lambda S_{R}} - 
   64 J^2 \pi^4 \lambda^4\Big)
   \end{equation}

and

\begin{equation}
B=\lambda \Big(-1 + 6 e^{\lambda S_{R}} - 14 e^{2 \lambda S_{R}} + 16 e^{3 \lambda S_{R}} - 
   9 e^{4 \lambda S_{R}} + 2 e^{5 \lambda S_{R}} + 64 J^2 \pi^4 \lambda^4 + 
   128 J^2 \pi^4 \lambda^4 e^{\lambda S_{R}} \Big)
\end{equation}

\begin{figure}[h]	
	\centering
	\begin{subfigure}{0.37\textwidth}
		\includegraphics[width=\linewidth]{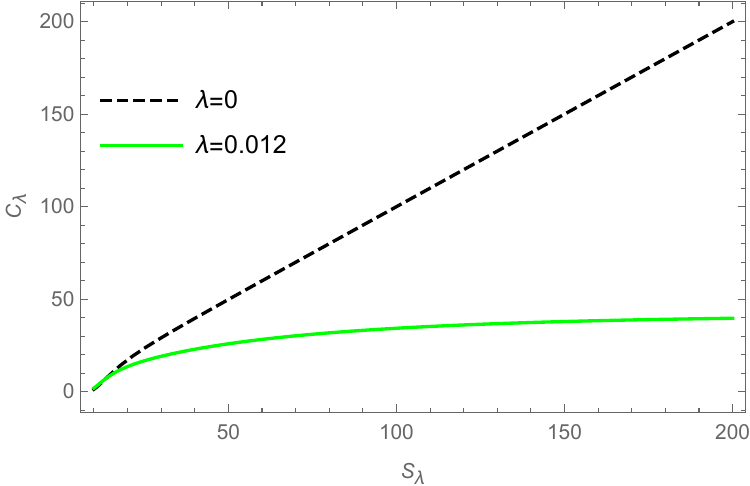}
		\caption{Heat capacity in the fixed $(J)$ ensemble for $l=1$ and $J=1$ }
		\label{5a}
		\end{subfigure}
		\begin{subfigure}{0.37\textwidth}
		\includegraphics[width=\linewidth]{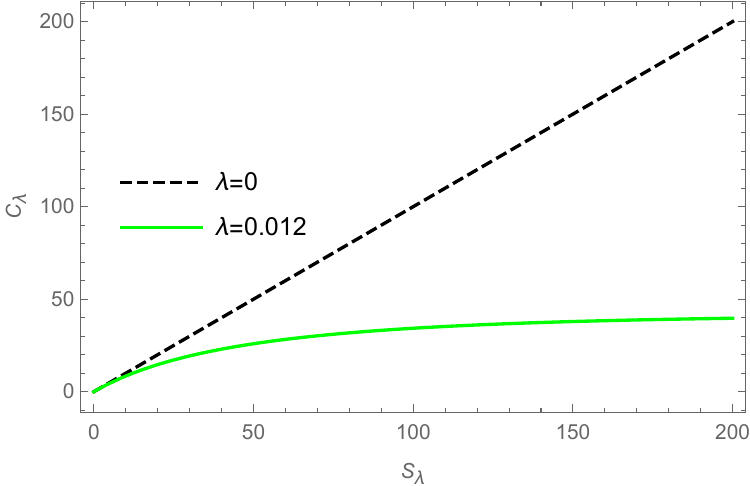}
		\caption{Heat capacity in the fixed $(\Omega)$ ensemble for $l=1$ and $\Omega=0.2$ }
		\label{5b}
		\end{subfigure}
		\caption{ The heat capacity versus Renyi entropy plots in the fixed $(J)$ and fixed $(\Omega)$ ensembles.}
	\label{5}
 \end{figure}

 We see from Fig.\ref{5a} that the heat capacity, $C_R$ for the  rotating BTZ black hole is plotted against the Renyi entropy, $S_R$ in the fixed $(J)$ ensemble. Here the solid green curve represents the heat capacity for the Renyi parameter, $\lambda=0.012$ whereas the black dashed line represents the heat capacity for the Bekenstein-Hawking(BH) entropy case $(\lambda=0)$. For the parameter $l=1$ and $ J=1$ we find that there are no divergences in the heat capacity curve for both the Renyi and Bekenstein-Hawking (BH) entropy cases. \\
 
 The  Gibbs free energy for the Renyi entropy case in the fixed $(J)$ ensemble can be obtained as:

$$G_{R} =   M_{R} - T_{R} S_{R}$$

\begin{equation}
G_{R} = 
\frac{\left(-1 + e^{\lambda S_{R}}\right) \left(64 J^2 \pi^4 \lambda^4 + \left(-1 + e^{\lambda S_{R}}\right)^4\right)  - 2 \lambda S_{R} e^{\lambda S_{R}} \left(1 - 4 e^{\lambda S_{R}} + 6 e^{2 \lambda S_{R}} - 4 e^{3 \lambda S_{R}} + e^{4 \lambda S_{R}} - 64 J^2 \pi^4 \lambda^4\right)}{16 \left(-1 + e^{\lambda S_{R}}\right)^3 \pi^2 \lambda^2}
\end{equation}\\

\begin{figure}[h]	
	\centering
	\begin{subfigure}{0.37\textwidth}
		\includegraphics[width=\linewidth]{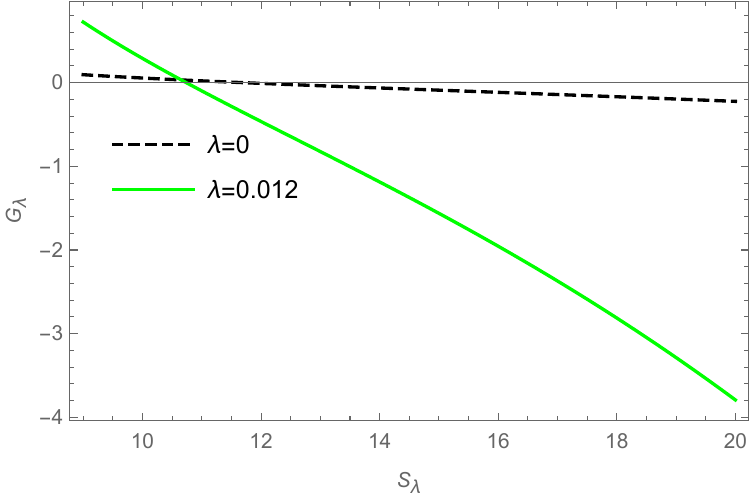}
		\caption{Free energy in the fixed $(J)$ ensemble for $l=1$ and $J=1$ }
		\label{6a}
		\end{subfigure}
		\begin{subfigure}{0.39\textwidth}
		\includegraphics[width=\linewidth]{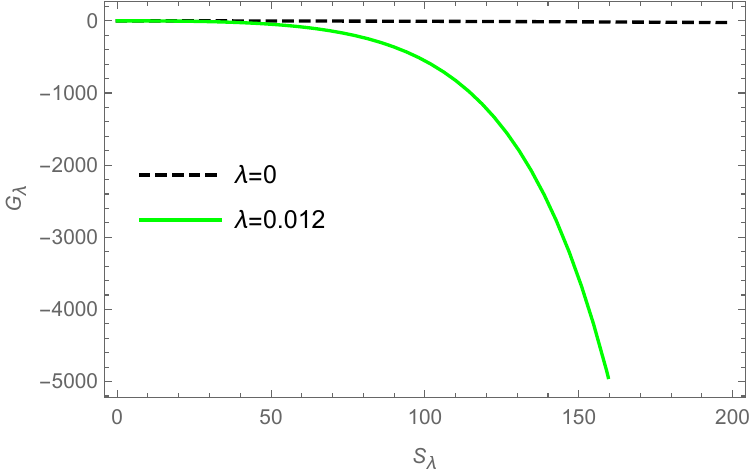}
		\caption{Free energy in  the fixed $(\Omega)$ ensemble for $l=1$ and $\Omega=0.2$.}
		\label{6b}
		\end{subfigure}
		\caption{ The Gibbs free energy versus Renyi entropy plots in the fixed $(J)$ and fixed $(\Omega)$ ensembles.}
	\label{6}
 \end{figure}

In Fig.\ref{6a} the Gibbs free energy, $G_R$ is plotted against the Renyi entropy, $S_R$ for $l=1$ and $J=1$ in the fixed $(J)$ ensemble. Here the solid green curve represents the Gibbs free energy for the Renyi parameter, $\lambda=0.012$ whereas the black dashed line represents the free energy for the Bekenstein-Hawking(BH) entropy case $(\lambda=0)$. The free energy for the $\lambda=0$ case has been scaled down by multiplying it with $\frac{1}{10}$  whereas the free energy for the $\lambda=0.012$ case is kept the same, this difference in scaling is just so that the distinct behaviour of both the curves could be seen more clearly.  We find that in Fig.\ref{6a} the Gibbs free energy goes to zero for both the entropy cases, where for $\lambda=0$, the zero point is at $S_R =11.69$ and for $\lambda=0.012$, the zero point is found to be at $S_R =10.74$, these zero points refer to the presence of a Hawking-Page phase transition present in the rotating BTZ black hole for both the Bekenstein-Hawking as well as the Renyi entropy cases.\\	

 \underline{\textbf{Fixed $(\Omega)$ ensemble}}:\\
 
   The modified mass for the Renyi entropy case in the fixed $(\Omega)$ ensemble is given by:
 $$M_{R} ' = M_{R} - J \frac{\partial M_{R}}{\partial J}$$
 $$M_{R} ' = M_{R} - J \Omega$$
 where $\frac{\partial M_{R}}{\partial J}= \Omega$ is the angular velocity of the black hole with respect to the angular momentum J. By making the necessary replacements and putting $l=1$, we get the final equation as follows:
\begin{equation}
M_R =  \frac{ ( 1 -\Omega^2) \left( e^{\lambda S_{R}} - 1\right)^2 }{16 \pi^2 \lambda^2 } 
\label{eqtn17}
\end{equation}

 We have here used the term $M_{R}$ instead of $M_{R}'$ for convenience. Here $M_{R}$ is the modified mass for the fixed $(\Omega)$ ensemble. The heat capacity for the Renyi entropy case in the fixed $(\Omega)$ ensemble can be calculated as:
$$C_{R}=  \frac{\frac{\partial M_{R}}{\partial S_{R}}}{\frac{\partial^2 M_{R}}{\partial S_{R} ^2}}\\=  \frac{ e^{\lambda S_R} - 1}{ \lambda \left( 2 e^{\lambda S_R} - 1\right) }$$\\

   The heat capacity of the rotating BTZ black hole is independent of $\Omega$ as can be seen from the above expression and therefore the presence or absence of Davies type phase transitions is independent of $\Omega$. We see from Fig.\ref{5b} that the heat capacity, $C_R$ for the  rotating BTZ black hole is plotted against the Renyi entropy, $S_R$ in the fixed $(\Omega)$ ensemble. Here the solid green curve represents the heat capacity for the Renyi parameter, $\lambda=0.012$ whereas the black dashed line represents the heat capacity for the Bekenstein-Hawking(BH) entropy case $(\lambda=0)$.  We see that for $l=1$ there are no divergences in the heat capacity curve for both the Renyi and Bekenstein-Hawking (BH) entropy cases.\\
    
    The  Gibbs free energy for the Renyi entropy case in the fixed $(\Omega)$ ensemble can be obtained as:

$$G_{R} =   M_{R} - T_{R} S_{R}$$

\begin{equation}
G_{R} = \frac{\left(-1 + e^{\lambda S_{R}}\right) \left(1 + e^{\lambda S_{R}} \left(-1 + 2 \lambda S_{R}\right)\right) \left(-1 + \Omega^2\right)}{16 \lambda^2 \pi^2}
\end{equation}\\

In Fig.\ref{6b} the Gibbs free energy, $G_R$ is plotted against the Renyi entropy, $S_R$ for $l=1$ and $\Omega=0.2$ in the fixed $(\Omega)$ ensemble. Here the solid green curve represents the Gibbs free energy for the Renyi parameter, $\lambda=0.012$ whereas the black dashed line represents the free energy for the Bekenstein-Hawking(BH) entropy case $(\lambda=0)$.  The free energy for the $\lambda=0$ case has been scaled down by multiplying it with $\frac{1}{10}$  whereas the free energy for the $\lambda=0.012$ case is kept the same, this difference in scaling is just so that the distinct behaviour of both the curves could be seen more clearly. We find that unlike the Kaniadakis entropy there is no Hawking-Page phase transition present in the Renyi entropy case for the fixed $(\Omega)$ ensemble.\\

 \subsection{The Barrow entropy case}
	The Barrow entropy, $S_{\Delta}$ in terms of the black hole entropy, $S_{BH}$  is given by (\ref{eqtn8}) as:
	 $$S_{\Delta}= (S_{BH})^{1 + \Delta/2}$$
  where, `$\Delta$' is the parameter that is linked to the  fractal structure of the system and for the limit $\Delta \rightarrow 0$, $S_{\Delta} \rightarrow S_{BH}$. Starting with the above equation we get the modified horizon radius for the R-BTZ black hole as:
  	$$S_{\Delta}= (4 \pi r_{+})^{1 + \Delta/2}$$

 \begin{equation}
 r_{+} = \frac{S_{\Delta}^{\frac{2}{2 + \Delta}}}{4 \pi}
 \label{eqtn24}
 \end{equation}
 \underline{\textbf{Fixed $(J)$ ensemble}}: \\
  We replace the horizon radius in  (\ref{eqtn9}) by  the one obtained in (\ref{eqtn24}) to get the modified mass $M_{\Delta}$ for the R-BTZ black hole which is given by :
$$M_{\Delta} =4 J^2 \pi^2 S_\Delta^{-\left(\frac{4}{2 + \Delta}\right)} +  \left(\frac{S_\Delta^{\left(\frac{4}{2 + \Delta}\right)}}{16 \pi^2}\right) 
$$\\
 The heat capacity for the Barrow entropy case in the fixed $(J)$ ensemble can be calculated as:
 $$C_{\Delta}=  \frac{\frac{\partial M_{\Delta}}{\partial S_{\Delta}}}{\frac{\partial^2 M_{\Delta}}{\partial S_{\Delta} ^2}}\\ = 
-\frac{S_\Delta \left(-64 J^2 \pi^4 + S_\Delta^{\frac{8}{2 + \Delta}}\right) (2 + \Delta)}{-384 J^2 \pi^4 - 2 S_\Delta^{\frac{8}{2 + \Delta}} - 64 J^2 \pi^4 \Delta + S_\Delta^{\frac{8}{2 + \Delta}} \Delta}$$\\

\begin{figure}[h]	
	\centering
	\begin{subfigure}{0.37\textwidth}
		\includegraphics[width=\linewidth]{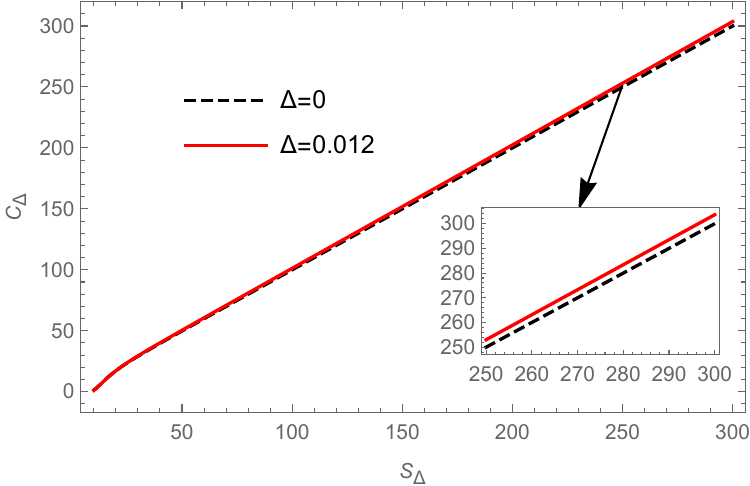}
		\caption{Heat capacity in the fixed $(J)$ ensemble for $l=1$ and $J=1$ }
		\label{7a}
		\end{subfigure}
		\begin{subfigure}{0.40\textwidth}
		\includegraphics[width=\linewidth]{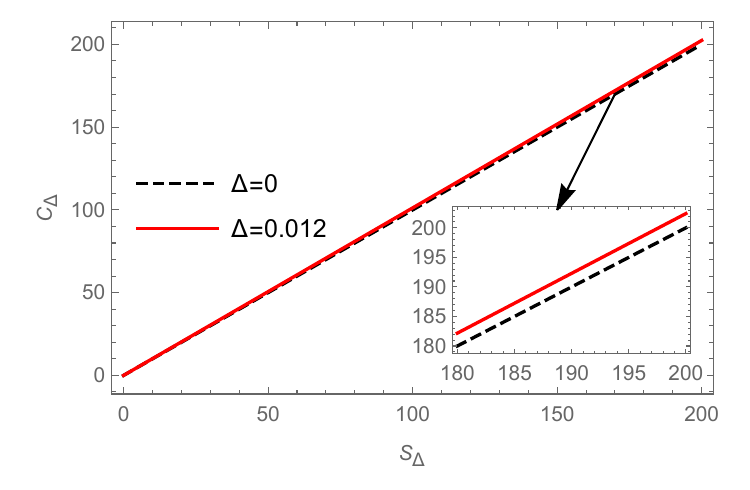}
		\caption{Heat capacity in the fixed $(\Omega)$ ensemble for $l=1$ and $\Omega=0.2$ }
		\label{7b}
		\end{subfigure}
		\caption{  The heat capacity versus Barrow entropy plots in the fixed $(J)$ and fixed $(\Omega)$ ensembles.}
	\label{7}
 \end{figure}

 We see from Fig.\ref{7a} that the heat capacity, $C_{\Delta}$ for the rotating BTZ black hole is plotted against the Barrow entropy, $S_{\Delta}$ in the fixed $(J)$ ensemble. Here the solid red curve represents the heat capacity for the Barrow parameter, $\Delta=0.012$ whereas the black dashed line represents the heat capacity for the Bekenstein-Hawking(BH) entropy case $(\Delta=0)$. For the parameter $l=1$ and $ J=1$ we find that there are no divergences in the heat capacity curve for both the Barrow and Bekenstein-Hawking (BH) entropy cases.\\
 
  The  Gibbs free energy for the Renyi entropy case in the fixed $(J)$ ensemble can be obtained as:

$$G_{\Delta} =   M_{\Delta} - T_{\Delta} S_{\Delta}$$

\begin{equation}
G_{\Delta} = 
\frac{S_{\Delta}^{-\frac{4}{2 + \Delta}} \left(S_{\Delta}^{\frac{8}{2 + \Delta}} (-2 + \Delta) + 64 J^2 \pi^4 (6 + \Delta)\right)}{16 \pi^2 (2 + \Delta)}
\end{equation}\\

\begin{figure}[h]	
	\centering
	\begin{subfigure}{0.37\textwidth}
		\includegraphics[width=\linewidth]{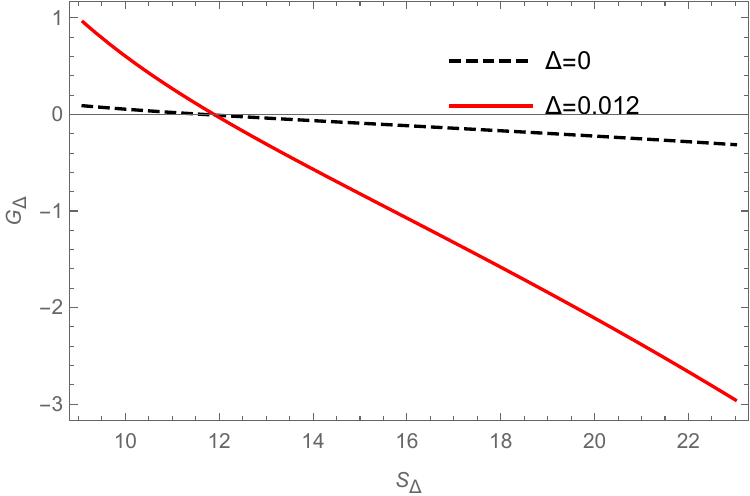}
		\caption{Free energy in the fixed $(J)$ ensemble for $l=1$ and $J=1$ }
		\label{8a}
		\end{subfigure}
		\begin{subfigure}{0.39\textwidth}
		\includegraphics[width=\linewidth]{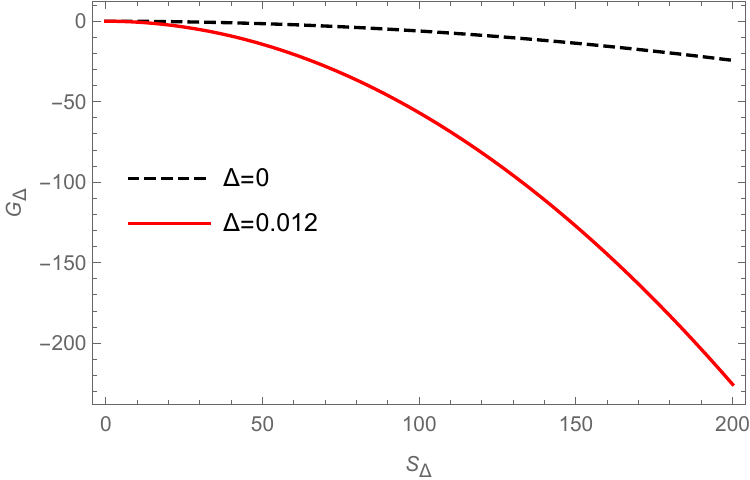}
		\caption{Free energy in  the fixed $(\Omega)$ ensemble for $l=1$ and $\Omega=0.2$.}
		\label{8b}
		\end{subfigure}
		\caption{ The Gibbs free energy versus Barrow entropy plots in the fixed $(J)$ and fixed $(\Omega)$ ensembles.}
	\label{8}
 \end{figure}

In Fig.\ref{8a} the Gibbs free energy, $G_\Delta$ is plotted against the Barrow entropy, $S_\Delta$ for $l=1$ and $J=1$ in the fixed $(J)$ ensemble. Here the solid red curve represents the Gibbs free energy for the Barrow parameter, $\Delta=0.012$ whereas the black dashed line represents the free energy for the Bekenstein-Hawking(BH) entropy case $(\Delta=0)$.  The free energy for the $\Delta=0$ case has been scaled down by multiplying it with $\frac{1}{10}$  whereas the free energy for the $\Delta=0.012$ case is kept the same, this difference in scaling is just so that the distinct behaviour of both the curves could be seen more clearly.We find that in Fig.\ref{8a} the Gibbs free energy goes to zero for both the entropy cases, where for $\Delta=0$, the zero point is at $S_R =11.69$ and for $\Delta=0.012$, the zero point is found to be at $S_R =11.89$, these zero points refer to the presence of a Hawking-Page phase transition present in the rotating BTZ black hole for both the Bekenstein-Hawking as well as the Barrow entropy cases.\\	
 
\underline{\textbf{Fixed $(\Omega)$ ensemble}}:\\

  The modified mass for the Barrow entropy case in the fixed $(\Omega)$ ensemble is given by:
 $$M_{\Delta} ' = M_{\Delta} - J \frac{\partial M_{\Delta}}{\partial J}$$
 $$M_{\Delta} ' = M_{\Delta} - J \Omega$$
 where $\frac{\partial M_{\Delta}}{\partial j}= \Omega$ is the angular velocity of the black hole with respect to the angular momentum, J. By making the necessary replacements and putting $l=1$, we get the final equation as follows:
 \begin{equation}	
M_{\Delta} = \frac{S_\Delta^{\frac{4}{2 + \Delta}} (1 - \Omega^2)}{16 \pi^2}
\label{eqtn26}
\end{equation}\\

 We have here used the term $M_{\Delta}$ instead of $M_{\Delta}'$ for convenience. Here $M_{\Delta}$ is the modified mass for the fixed $(\Omega)$ ensemble. The heat capacity for the Barrow entropy case in the fixed $(\Omega)$ ensemble can be calculated as:
 $$C_{\Delta}=  \frac{\frac{\partial M_{\Delta}}{\partial S_{\Delta}}}{\frac{\partial^2 M_{\Delta}}{\partial S_{\Delta} ^2}}\\= 
\frac{(2 + \Delta) S_\Delta }{(2 - \Delta) }
 $$\\

The heat capacity of the rotating BTZ black hole is independent of $\Omega$ as can be seen from the above expression and therefore the presence or absence of Davies type phase transitions is independent of $\Omega$. We see from Fig.\ref{7b} that the heat capacity, $C_{\Delta}$ for the  rotating BTZ black hole is plotted against the Barrow entropy, $S_{\Delta}$ in the fixed $(\Omega)$ ensemble. Here the solid red curve represents the heat capacity for the Barrow parameter, $\Delta=0.012$ whereas the black dashed line represents the free energy for the Bekenstein-Hawking(BH) entropy case $(\Delta=0)$. We see that for $l=1$  there are no divergences in the heat capacity curve for both the Barrow and Bekenstein-Hawking (BH) entropy cases.\\

 The  Gibbs free energy for the Barrow entropy case in the fixed $(\Omega)$ ensemble can be obtained as:

$$G_{\Delta} =   M_{\Delta} - T_{\Delta} S_{\Delta}$$

\begin{equation}
G_{\Delta} = \frac{S_{\Delta}^{\frac{4}{2 + \Delta}} (2 - \Delta) (-1 + \Omega^2)}{16 \pi^2 (2 + \Delta)}
\end{equation}\\

In Fig.\ref{8b} the Gibbs free energy, $G_\Delta$ is plotted against the Barrow entropy, $S_\Delta$ for $l=1$ and $\Omega=0.2$ in the fixed $(\Omega)$ ensemble. Here the solid red curve represents the Gibbs free energy for the Renyi parameter, $\Delta=0.012$ whereas the black dashed line represents the free energy for the Bekenstein-Hawking(BH) entropy case $(\Delta=0)$. Here the solid red curve represents the Gibbs free energy for the Barrow parameter, $\Delta=0.012$ whereas the black dashed line represents the free energy for the Bekenstein-Hawking(BH) entropy case $(\Delta=0)$. The free energy for the $\Delta=0$ case has been scaled down by multiplying it with $\frac{1}{10}$  whereas the free energy for the $\Delta=0.012$ case is kept the same, this difference in scaling is just so that the distinct behaviour of both the curves could be seen more clearly.  We find that unlike the Kaniadakis entropy there is no Hawking-Page phase transition present in the Barrow entropy case for the fixed $(\Omega)$ ensemble.\\	

 
 \section{Thermodynamic geometry of the R-BTZ black hole}

We discuss the thermodynamic geometry  of the rotating BTZ black hole for both the Ruppeiner and geometrothermodynamic (GTD) formalisms. The Ruppeiner metric is defined as the negative hessian of entropy with respect to other extensive variables. Therefore it is only the fixed $(J)$ ensemble which holds good in the Ruppeiner formalism as  $J$ is an extensive thermodynamic quantity whereas $\Omega$ is not.\\

 Therefore the fixed $(\Omega)$ ensemble do not hold good as per the definition of the Ruppeiner metric. We therefore discuss the thermodynamic geometry of  the fixed $(\Omega)$ ensemble in the GTD formalism which is defined on a multi-dimensional phase space that comprises of both the extensive and intensive  variables of the thermodynamic system. The GTD formalism is therefore the appropriate geometric formalism for discussing both the thermodynamic ensembles.\\
 
 We first discuss the Ruppeiner geometry of the rotating BTZ black hole in the fixed $(J)$ ensemble where we observe  a curvature singularity in the Ruppeiner scalar curve only for the Kaniadakis entropy case. We then discuss the GTD geometry of the black hole for both the thermodynamic ensembles, where we observe curvature singularities in the GTD scalar curves in Kaniadakis entropy case only. These curvature singularities are seen to occur exactly at the points where we saw divergences in the corresponding heat capacity curves in the previous section.
 
\subsection{\textbf{Ruppeiner formalism}}
 The Ruppeiner metric for the R-BTZ black hole in the fixed $(J)$ ensemble  can be obtained from the relation (\ref{eqtn2}) and is given as :-
$$dS_{R}^{2}= \frac{1}{T} \left(\frac{\partial^2 M}{\partial S^2}dS^2 + \frac{\partial^2 M}{\partial J^2} dJ^2  + 2\frac{\partial^2 M}{\partial S \partial J} dS dJ\right) $$
 where  $M$, $S$, $T$ and $J$ are the mass, entropy, Hawking Temperature and angular momentum for the rotating BTZ black hole respectively.\\
 
From the above metric we calculate the Ruppeiner scalar for all the non-extensive entropy cases in the fixed $(J)$ ensemble. We find that there is a curvature singularity in the Ruppeiner scalar curve \textbf{only for the Kaniadakis entropy case}. For all the other entropy cases namely: the Renyi and Barrow entropy cases the Ruppeiner scalar curves remain regular with no singularities therein.  We have not presented here either the derivation or the final expressions for the Ruppeiner scalar due to their considerable length. However obtaining these expressions is fairly straight forward and involves only routine mathematical computations. We present a detailed analysis of the Ruppeiner thermodynamic geometry for the Kaniadakis entropy case in the  fixed $(J)$ ensemble as follows:\\ 

\begin{figure}[h!t]
  \centering
		\includegraphics[width=9cm,height=6cm]{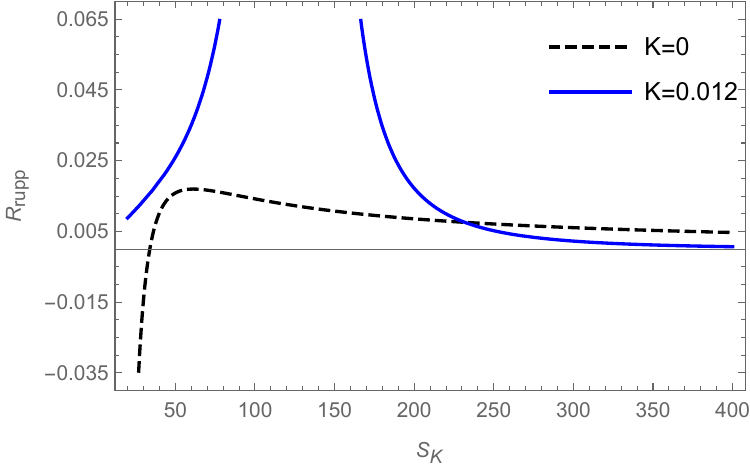}
		\caption{The Ruppeiner scalar versus Kaniadakis entropy plot in the fixed $(J)$ ensemble for $l=1$ and $J=1$.}
		\label{9}
		\end{figure}

We see from Fig.\ref{9} that the Ruppeiner scalar $R_{rupp}$ is plotted against the Kaniadakis entropy $S_K$ in the fixed $(J)$ ensemble.  We see that for $l=1$ and $J=1$, for the Kaniadakis parameter $K=0.012$ the Ruppeiner scalar curve has  a curvature singularity specifically at $S_K = 125.74$, which is depicted by the solid blue curve in the figure. The point of singularity here is dissimilar to the point of divergence $S_K=125.77$ obtained in the corresponding heat capacity curve for the Kaniadakis entropy case  in the fixed $(J)$ ensemble. For the $K=0$ case however the Ruppeiner scalar curve is found to be regular everywhere with no curvature singularities as is depicted in the figure by the black dashed curve.     
 \\
		
\subsection{\textbf{Geometrothermodynamic(GTD) formalism}}
To describe the  rotating BTZ black hole in GTD formalism, we first assume a five-dimensional phase space $\mathcal{T}$ with co-ordinates $M, S, J, T$ and $\Omega$  which are the mass, entropy, angular momentum,  temperature and angular velocity  for the black hole. The contact 1-form for these set of co-ordinates can be written as:-
$$\Theta = dM -TdS -\Omega dJ $$
The 1-form satisfies the condition $\Theta \wedge (d\Theta)^3 \neq 0$ and a legendre invariant metric that goes by :-
$$G = (dM -TdS -\Omega dJ )^2 + TS(-dT dS + d\Omega dJ )$$
We then assume a two-dimensional sub-space of $\mathcal{T}$ with co-ordinates $E^a$ where $a=1,2$. We denote this subspace with $\epsilon$ which is defined as a smooth mapping, $\phi: \epsilon \rightarrow \mathcal{T}$. The subspace $\epsilon$ becomes the space of equilibrium states if the pullback of $\phi$ is 0 i.e. $\phi^{*}(\Theta)=0$. A metric structure $g$ is then naturally induced on $\epsilon$ by applying the same pullback on the metric G of $\mathcal{T}$. This induced metric determines all the geometric properties of the equilibrium space for the R-BTZ black hole.
\subsubsection{\textbf{The Kaniadakis entropy case}}
\underline{{\textbf{Fixed $(J)$ ensemble}:}}\\

We write the GTD metric for the Kaniadakis entropy case in the fixed $(J)$ ensemble from the general metric given in  (\ref{eqtn3}). We consider the thermodynamic potential $\varphi$ to be the mass $M_K$ for the Kaniadakis entropy case in  the fixed $(J)$ ensemble as obtained in equation (\ref{eqtn11}). The GTD metric is given by :- 
$$g  =S_K \left(\frac{\partial M_K}{\partial S_K}\right)\left(- \frac{\partial^2 M_K}{\partial S_K^2} dS_K^2 + \frac{\partial^2 M_K}{\partial J^2} dJ^2 \right) $$  
\\

From the above metric we calculate the GTD scalar for  the Kaniadakis entropy case in the fixed $(J)$ ensemble. We have not presented here either the derivation or the final expressions for the GTD scalar due to their considerable length. However obtaining these expressions is fairly straight forward and involves only routine mathematical computations. We present a detailed analysis of the GTD thermodynamic geometry for the Kaniadakis entropy case in the  fixed $(J)$ ensemble as follows:\\

\begin{figure}[h!t]
  \centering
		\includegraphics[width=9cm,height=6cm]{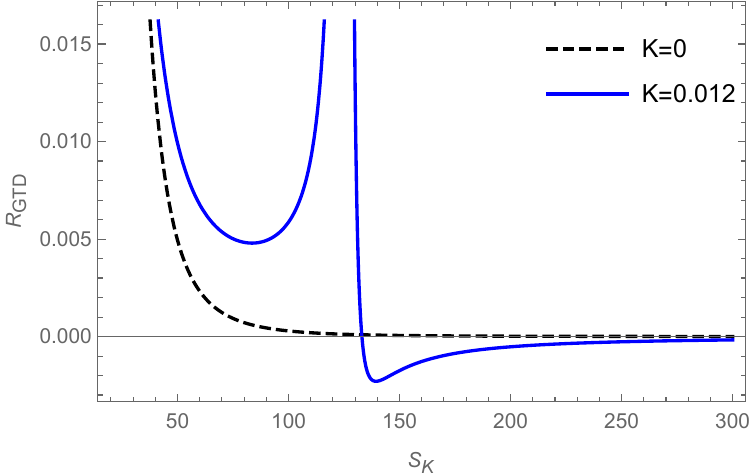}
		\caption{The GTD scalar versus Kaniadakis entropy plot in the fixed $(J)$ ensemble for $l=1$ and $J=1$.}
		\label{10}
		\end{figure}

We see from Fig.\ref{10} that the GTD scalar $R_{GTD}$ is plotted against the Kaniadakis entropy $S_K$  for the Kaniadakis parameter $K=0.012$ in the fixed $(J)$ ensemble. We find that for  $l=1$ and $J=1$, the GTD scalar has a curvature singularity at $S_K = 125.77$ for $K=0.012$ as depicted by the blue solid curve whereas for the BH entropy case ($K=0$) the curve remains regular  everywhere with no curvature singularities, as is depicted in the figure by the black dashed curve. Unlike Ruppeiner scalar, the singularity obtained from the GTD scalar curve matches exactly with the point of divergence obtained from the corresponding heat capacity curve for the Kaniadakis entropy case in the fixed $(J)$ ensemble.
 \\
		
		\underline{{\textbf{Fixed $(\Omega)$ ensemble}:}}\\

We write the GTD metric for the Kaniadakis entropy case in the fixed $(\Omega)$ ensemble from the general metric given in  (\ref{eqtn3}). We consider the thermodynamic potential $\varphi$ to be the mass $M_K$ for the Kaniadakis entropy case in  the fixed $(\Omega)$ ensemble as obtained in equation (\ref{eqtn14}). The GTD metric is given by :- 
$$g  =S_K \left(\frac{\partial M_K}{\partial S_K}\right)\left(- \frac{\partial^2 M_K}{\partial S_K^2} dS_K^2 + \frac{\partial^2 M_K}{\partial \Omega^2} d\Omega^2 \right) $$  
\\

From the above metric we calculate the GTD scalar for  the Kaniadakis entropy case in the fixed $(\Omega)$ ensemble. We have not shown here either the derivation or the final expressions for the GTD scalar due to their considerable length. However obtaining them is fairly straight forward and involves only routine mathematical computations. We present a detailed analysis of the GTD thermodynamic geometry for the Kaniadakis entropy case in the  fixed $(\Omega)$ ensemble as follows:\\

\begin{figure}[h!t]
  \centering
		\includegraphics[width=9cm,height=6cm]{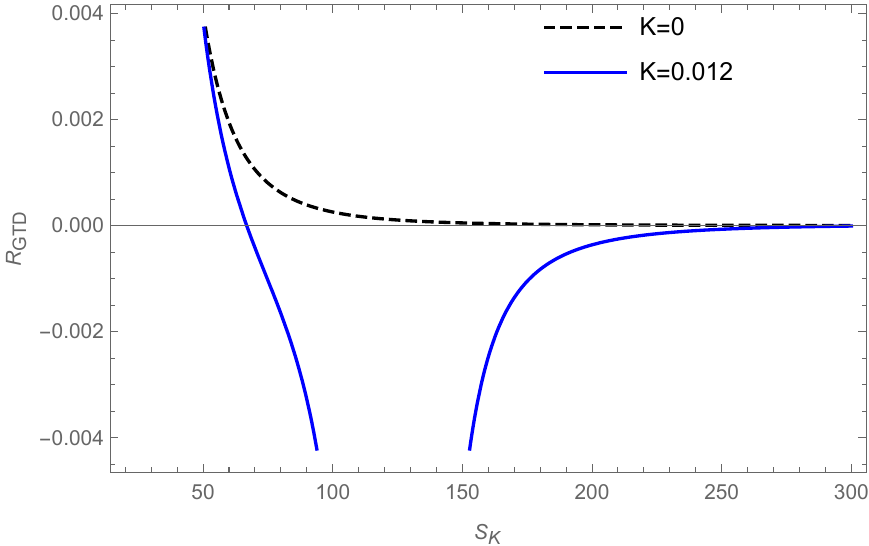}
		\caption{The GTD scalar versus Kaniadakis entropy plot in  the fixed $(\Omega)$ ensemble  for $l=1$ and $\Omega=0.2$}
		\label{11}
		\end{figure}

We see from Fig.\ref{11} that the GTD scalar $R_{GTD}$ is plotted against the Kaniadakis entropy $S_{K}$  in the fixed $(\Omega)$ ensemble. We find that for  $l=1$ and $\Omega=0.2$, the GTD scalar has a curvature singularity at $S_K =125.74$ for $K=0.012$ as depicted by the blue solid curve whereas for the BH entropy case ($K=0$) the curve remains regular  everywhere with no curvature singularities, as is depicted in the figure by the black dashed curve. Unlike Ruppeiner scalar, the singularity obtained from the GTD scalar curve matches exactly with the point of divergence obtained from the corresponding heat capacity curve for the Kaniadakis entropy case in the fixed $(\Omega)$ ensemble.

		\section{Thermodynamics of the charged BTZ black hole}
  The action which facilitates the field equations\cite{banados1992black,clement1993classical,
clement1996spinning,maeda2024charged,konewko2024charge}
from which the $(2+1)$ dimensional charged BTZ black hole solutions are obtained is given as:
$$I=\frac{1}{2\pi} \int d^3 x \sqrt{-g}\biggl(R - 2\Lambda - \frac{\pi}{2}F_{\mu \nu }F^{\mu \nu }\biggr)$$
where, the AdS length $l$ is related to the cosmological constant $\Lambda$ by the relation: $\Lambda=- \frac{1}{l^2}$ and the Einstein's field equations are given by:
$$G_{\mu \nu} - \Lambda g_{\mu \nu}= \pi T_{\mu \nu}$$
and the energy momentum tensor for the same is given as:
$$T_{\mu \nu}= F_{\mu \rho} F_{\nu \sigma} g^{\rho \sigma} -\frac{1}{4}g_{\mu \nu}F^2$$\\

The line element that corresponds to the given solution is:
$$ds^2 = -f(r) dt^2 + \frac{dr^2}{f(r)} + r^2 d\phi^2$$
where $f(r) = -M + \frac{r^2}{l^2}  - \frac{\pi Q^2}{2} \ln \left[r\right]$ and `M' and `Q'  are the mass and electric charge  carried by the black hole. Solving the function for $f(r)=0$ would give us the roots which determine the horizon radius. For an exterior horizon radius, `$r_{+}$' the black hole mass is given by:
\begin{equation}
M=\frac{r_{+}^2}{l^2} +  \frac{\pi Q^2}{2} \ln \left[r_{+}\right]
\label{eqtn27}
\end{equation}
and the black hole entropy is given by:
$$S_{BH}= 4 \pi r_{+}$$
\subsection{The Kaniadakis entropy case}

  \underline{\textbf{Fixed $(Q)$ ensemble}}: \\
 
 We replace the horizon radius in  (\ref{eqtn27}) by  the one obtained in (\ref{eqtn10}) to get the modified mass $M_{K}$ for the charged BTZ black hole which is given by :
$$M_{K}= \frac{ArcSinh^2 [K S_{K}]}{16 \pi^2 l^2 K^2} - \frac{\pi Q^2}{2} \ln \biggl[\frac{ArcSinh[K S_{K}]}{4 \pi K}\biggr] $$
 The function, $ArcSinh[K S_{K}]$ can be replaced by its logarithmic form, $\ln \biggl[KS_{K}+ \sqrt{1 + K^2 S_{K} ^2}\biggr]$ and therefore the expression for mass becomes:
 \begin{equation}
M_{K}=  \frac{\ln [KS_{K}+ \sqrt{1 + K^2 S_{K} ^2}]^2}{16 \pi^2 l^2 K^2} - \frac{\pi Q^2}{2} \ln \biggl[\frac{\ln [KS_{K}+ \sqrt{1 + K^2 S_{K} ^2}]}{4 \pi K}\biggr]
\label{eqtn28}
\end{equation}\\
 
The  heat capacity for the Kaniadakis entropy case in the fixed $(Q)$ ensemble can be obtained as:
\begin{equation}
C_{K} =  \frac{\frac{\partial M_{K}}{\partial S_{K}}}{\frac{\partial^2 M_{K}}{\partial S_{K} ^2}}= \frac{A}{B}
\end{equation}
where,
\begin{equation}
A=  \left( 1 + K^2 S_K^2 \right) \ln \left( K S_K + \sqrt{1 + K^2 S_K^2} \right) \left( -4 K^2 \pi^3 Q^2 + \ln \left( K S_K + \sqrt{1 + K^2 S_K^2} \right)^2 \right)
\end{equation}
and
\begin{align}
B = K \biggl( 
    & 4 K^3 \pi^3 Q^2 
      \ln \biggl( K S_K + \sqrt{1 + K^2 S_K^2} \biggr) 
     - K \ln \biggl( K S_K + \sqrt{1 + K^2 S_K^2} \biggr)^3 S_K  + \biggl( 4 K^2 \pi^3 Q^2\notag \\
    & + \ln \biggl( K S_K + \sqrt{1 + K^2 S_K^2} \biggr)^2 \biggr) 
      \sqrt{1 + K^2 S_K^2} 
\biggr).
\end{align}

In Fig.\ref{12} the heat capacity, $C_K$ is plotted against the Kaniadakis entropy, $S_K$ for $l=1$ and $J=1$ in the fixed $(Q)$ ensemble. Here the solid blue curve represents the heat capacity for the Kaniadakis parameter, $K=0.012$ whereas the black dashed line represents the heat capacity for the Bekenstein-Hawking(BH) entropy case $(K=0)$. We find that for $K=0.012$, the heat capacity has a divergence at $S_K=128.82$ which indicates  the presence of a Davies type phase transition present in the Kaniadakis modified black hole whereas the heat capacity for the  BH entropy case shows no such behaviour.	
\begin{figure}[h!t]
		\centering
		\includegraphics[width=9cm,height=6cm]{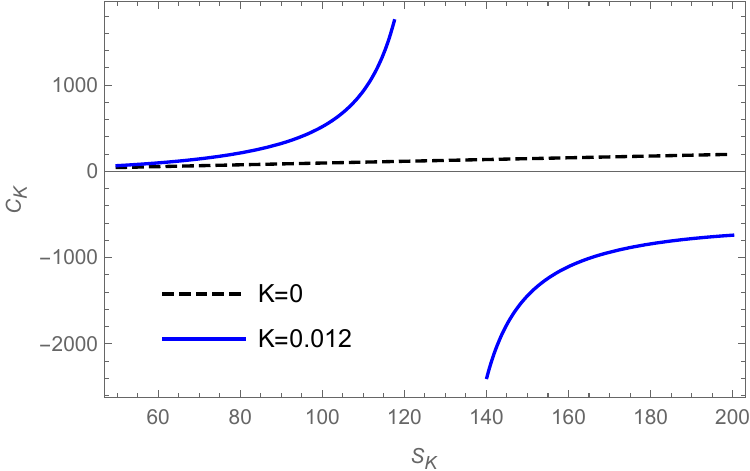}
		\caption{The heat capacity versus Kaniadakis entropy plot in the fixed $(Q)$ ensemble for $l=1$ and $Q=1$}
		\label{12}
		\end{figure}\\ 
		
		The  Gibbs free energy for the Kaniadakis entropy case in the fixed $(Q)$ ensemble can be obtained as:

$$G_{K} =   M_{K} - T_{K} S_{K}$$

\begin{align}
G_K = 
&\ \frac{1}{2} \pi Q^2 \ln \left( 4 \pi \right) 
   - \frac{1}{2} \pi Q^2 \ln \left( \frac{\ln \bigl( K S_K + \sqrt{1 + K^2 S_K^2} \bigr)}{K} \right) + \frac{\ln \bigl( K S_K + \sqrt{1 + K^2 S_K^2} \bigr)^2}{16 \pi^2 K^2 }\notag \\
&\ + \frac{\pi K Q^2 S_K}{2  \sqrt{1 + K^2 S_K^2} \ln \bigl( K S_K + \sqrt{1 + K^2 S_K^2} \bigr)} - \frac{ S_K \ln \bigl( K S_K + \sqrt{1 + K^2 S_K^2} \bigr)}{8 \pi^2 K  \sqrt{1 + K^2 S_K^2}}.
\end{align}\\

\begin{figure}[h]	
	\centering
	\begin{subfigure}{0.37\textwidth}
		\includegraphics[width=\linewidth]{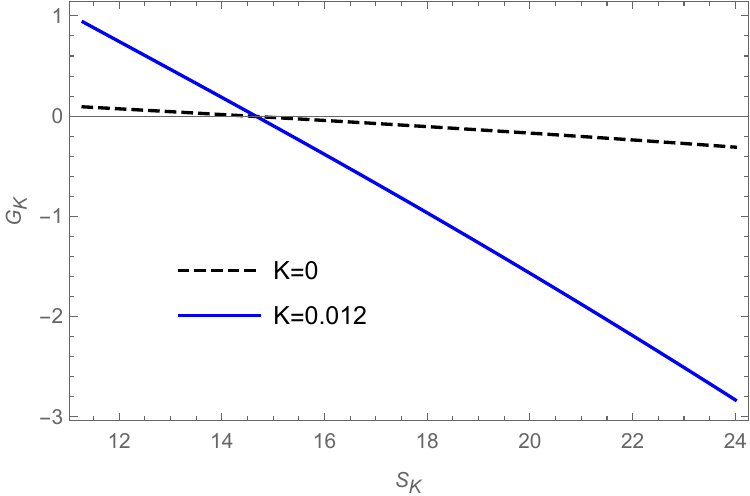}
		\caption{Gibbs free energy versus Kaniadakis entropy $(11< S_K <24)$.}
		\label{13a}
		\end{subfigure}
		\begin{subfigure}{0.37\textwidth}
		\includegraphics[width=\linewidth]{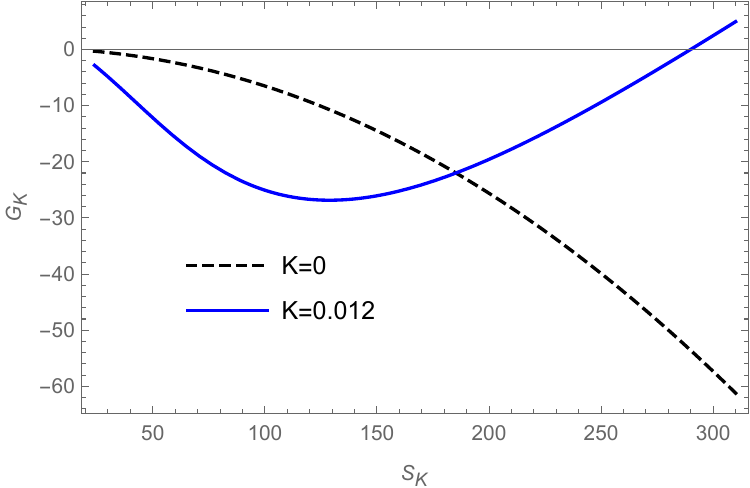}
		\caption{Gibbs free energy versus Kaniadakis entropy $(S_K > 24)$.}
		\label{13b}
		\end{subfigure}
		\begin{subfigure}{0.40\textwidth}
		\includegraphics[width=\linewidth]{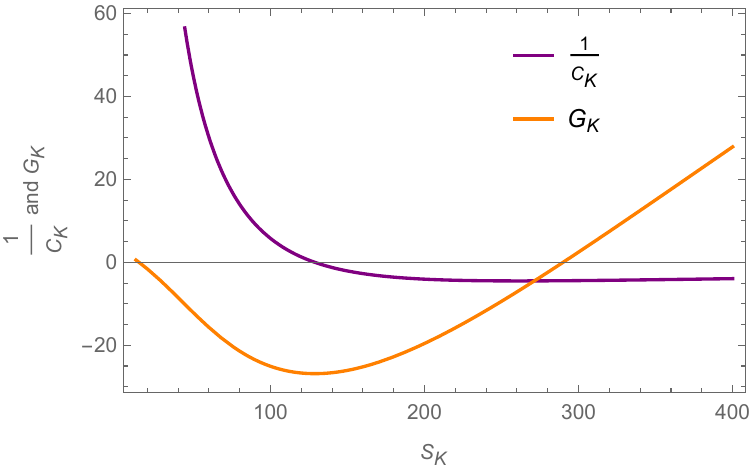}
		\caption{ Inverse specific heat and Gibbs free energy versus Kaniadakis entropy.}
		\label{13c}
		\end{subfigure}
		\caption{The Gibbs free energy curves for Kaniadakis entropy in  the fixed $(Q)$ ensemble for $l=1$ and $Q=1$.}
	\label{13}
 \end{figure}

In Fig.\ref{13a} and Fig.\ref{13b} the Gibbs free energy, $G_K$ is plotted against the Kaniadakis entropy, $S_K$ for $l=1$ and $Q=1$ in the fixed $(Q)$ ensemble. We draw two different plots so as to make the presence of the phase transitions more transparent. Here the solid blue curve represents the Gibbs free energy for the Kaniadakis parameter, $K=0.012$ whereas the black dashed line represents the free energy for the Bekenstein-Hawking(BH) entropy case $(K=0)$. The free energy for the $K=0$ case has been scaled down by multiplying it with $\frac{1}{10}$  whereas the free energy for the $K=0.012$ case is kept the same, this difference in scaling is just so that the distinct behaviour of both the curves could be seen more clearly. We keep in mind that the scaling does not affect the position or occurrence of the phase transitions. We find that in Fig.\ref{13a} the Gibbs free energy goes to zero for both the entropy cases, where for $K=0$, the zero point is at $S_K =14.55$ and for $K=0.012$, the zero point is found to be at $S_K =14.67$, these zero points refer to the presence of a Hawking-Page phase transition present in the rotating BTZ black hole for both the Bekenstein-Hawking as well as the Kaniadakis entropy cases. We again find that for $K=0.012$, the Gibbs free energy in Fig.\ref{13b} has a zero point at $S_K=289.97$ only for the Kaniadakis entropy case which indicates  the presence of a Hawking-Page phase transition present in the Kaniadakis modified black hole whereas the Gibbs free energy for the  BH entropy case shows no such behaviour. In  Fig.\ref{13c} we plot the inverse of specific heat capacity (purple) and the Gibbs free energy (orange) against the Kaniadakis entropy. The inverse of specific heat capacity  has been scaled up by multiplying it with $3000$  whereas the free energy is kept the same, this difference in scaling is just so that the distinct behaviour of both the curves could be seen more clearly. We keep in mind that the scaling does not affect the position or occurrence of the phase transitions. The points where these curves intersect the horizontal axis would represent the Davies-type (Purple) and Hawking-Page (Orange) phase transition respectively which are seen to be quite distant from each other.\\

 \underline{\textbf{Fixed $(\Phi)$ ensemble}}: \\
 
  The modified mass for the Kaniadakis entropy case in the fixed $(\Phi)$ ensemble is given by:
 $$M_{K} ' = M_{K} - Q \frac{\partial M_{K}}{\partial Q}$$
 where $\frac{\partial M_{K}}{\partial Q}= \Phi$ is the electric potential of the black hole with respect to the electric charge $Q$. By writing the term $M_{K} '$ as $M_{K}$ for convenience we get:
 \begin{equation}
 M_K = \frac{ \Phi^2}{ 2 \pi \left(\ln \left( \ln ( K S_K + \sqrt{1 + K^2 S_K^2 )}- \ln(4 \pi K )\right) \right)} + \frac{\ln \left( K S_K + \sqrt{1 + K^2 S_K^2} \right)^2} {16 \pi^2 K^2}
\end{equation}\\

The  heat capacity for the Kaniadakis entropy case in the fixed $(\Phi)$ ensemble can be obtained as:\\

  $$C_{K} =  \frac{\frac{\partial M_{K}}{\partial S_{K}}}{\frac{\partial^2 M_{K}}{\partial S_{K} ^2}}\\= \frac{A}{B}$$
where,
\begin{align}
A = &\ \biggl( 
    \ln(4 \pi) 
    - \ln \biggl( \frac{\ln \bigl( K S_K + \sqrt{1 + K^2 S_K^2} \bigr)}{K} \biggr) 
    \biggr) 
    \ln \bigl( K S_K + \sqrt{1 + K^2 S_K^2} \bigr)   \biggl( 
    -4 \pi K^2 \Phi^2\notag \\
&  + \ln \biggl( \frac{\ln \bigl( K S_K + \sqrt{1 + K^2 S_K^2} \bigr)}{4 K \pi} \biggr)^2 
    \ln \bigl( K S_K + \sqrt{1 + K^2 S_K^2} \bigr)^2 
    \biggr) 
    \bigl( 1 + K^2 S_K^2 \bigr)
\end{align}

and
\begin{align}
B = K \biggl( 
    & K \biggl( 
        \biggl( \ln \bigl( 4 \pi \bigr) 
        - \ln \biggl( \frac{\ln \bigl( K S_K + \sqrt{1 + K^2 S_K^2} \bigr)}{K} \biggr) \biggr) 
        \ln \bigl( K S_K + \sqrt{1 + K^2 S_K^2} \bigr) \biggr) 
     \biggl( 4 K^2 \pi \Phi^2 \notag \\  
      & - \ln \biggl( \frac{\ln \bigl( K S_K + \sqrt{1 + K^2 S_K^2} \bigr)}{4 K \pi} \biggr)^2\ln \bigl( K S_K + \sqrt{1 + K^2 S_K^2} \bigr)^2 
        \biggr) S_K + \biggl( 
        4 K^2 \pi \Phi^2 \bigl( -2 + \ln \bigl( 4 \pi \bigr) \bigr)\notag \\
    &  + \ln \bigl( 4 \pi \bigr)^3 
          \ln \bigl( K S_K + \sqrt{1 + K^2 S_K^2} \bigr)^2  \quad + 3 \ln \bigl( 4 \pi \bigr) 
          \ln \biggl( \frac{\ln \bigl( K S_K + \sqrt{1 + K^2 S_K^2} \bigr)}{K} \biggr)^2\notag \\
    &  \ln \bigl( K S_K + \sqrt{1 + K^2 S_K^2} \bigr)^2  \quad - \ln \biggl( \frac{\ln \bigl( K S_K + \sqrt{1 + K^2 S_K^2} \bigr)}{K} \biggr)^3 
          \ln \bigl( K S_K + \sqrt{1 + K^2 S_K^2} \bigr)^2 \notag \\
    & \quad - \ln \biggl( \frac{\ln \bigl( K S_K + \sqrt{1 + K^2 S_K^2} \bigr)}{K} \biggr) 
          \biggl( 
            4 K^2 \pi \Phi^2 
            + 3 \ln \bigl( 4 \pi \bigr)^2 
              \ln \bigl( K S_K + \sqrt{1 + K^2 S_K^2} \bigr)^2 
          \biggr) 
    \biggr) \sqrt{1 + K^2 S_K^2} \biggr)
\end{align}

\begin{figure}[h]	
	\centering
	\begin{subfigure}{0.37\textwidth}
		\includegraphics[width=\linewidth]{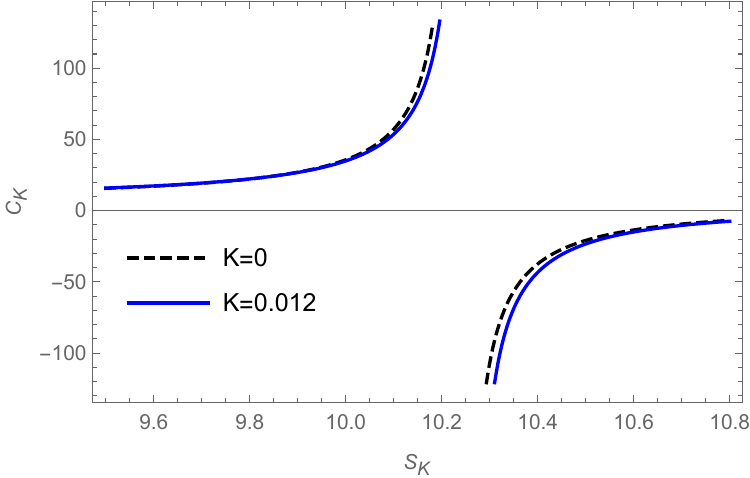}
		\caption{Heat capacity versus Kaniadakis entropy $(S < 11)$.}
		\label{14a}
		\end{subfigure}
		\begin{subfigure}{0.40\textwidth}
		\includegraphics[width=\linewidth]{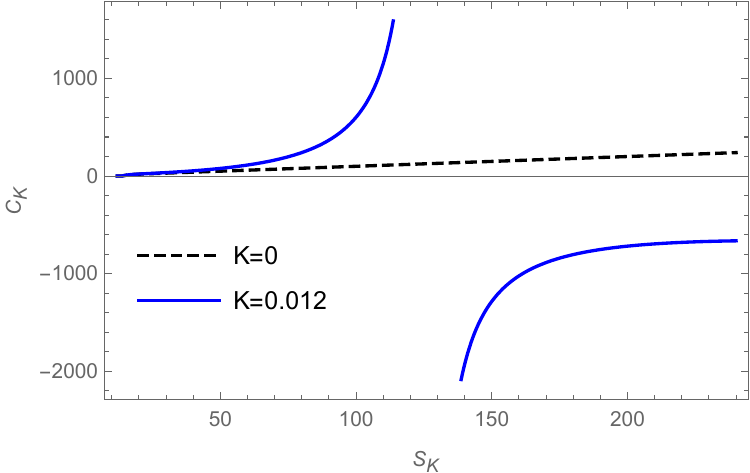}
		\caption{ Heat capacity versus Kaniadakis entropy $(11<S < 300)$.}
		\label{14b}
		\end{subfigure}
		\caption{The heat capacity versus Kaniadakis entropy plot in the fixed $(\Phi)$ ensemble for $l=1$ and $\Phi=0.2$}
	\label{14}
 \end{figure}

 In Fig.\ref{14} the heat capacity, $C_K$ for the charged  BTZ black hole is plotted against the Kaniadakis entropy, $S_K$ in the fixed $(\Phi)$ ensemble. We draw the same plot in two different entropy ranges so as to make all the appearing divergences in the heat capacity curve visible. We see from  Fig.\ref{14a} that for $l=1$ and $\Phi=0.2$ there is a divergence at $S_K=10.26$ for $K=0.012$ as shown by the solid blue curve. The heat capacity in the Bekenstein-Hawking(BH) entropy case ($K=0)$ here also produces a divergence at $S_K=10.24$  as shown by the black dashed curve. In Fig.\ref{14b} we see that there is a divergence in the heat capacity curve at $S_K=125.74$ for $K=0.012$ whereas here the heat capacity in the BH entropy case $(K=0)$ remains a monotonically increasing function of entropy with no divergences.	
 \\ 
		
		The  Gibbs free energy for the Kaniadakis entropy case in the fixed $(\Phi)$ ensemble can be obtained as:

$$G_{K} =   M_{K} - T_{K} S_{K}= \frac{\mathcal{C}}{\mathcal{D}}$$  

where,
\begin{align}
\mathcal{C} = &\ \frac{-\Phi^2 }{4 \pi \bigl( \ln(4 \pi) - \ln \bigl( \frac{\ln \bigl( K S_K + \sqrt{1 + K^2 S_K^2} \bigr)}{K} \bigr) \bigr)} + \frac{\ln \bigl( K S_K + \sqrt{1 + K^2 S_K^2} \bigr)^2}{16 K^2 \pi^2}  \notag \\
&\ - S_K \biggl( 
    -4 \pi K^2 \Phi^2 \ln \bigl( K S_K + \sqrt{1 + K^2 S_K^2} \bigr)^2  + \ln(4 \pi)^2 \ln \bigl( K S_K + \sqrt{1 + K^2 S_K^2} \bigr)^4 \notag \\
&\ \ \ \ \biggl(  \ln \bigl( \frac{\ln \bigl( K S_K + \sqrt{1 + K^2 S_K^2} \bigr)}{K} \bigr)-2 \ln(4 \pi)\biggr) \ln \bigl( \frac{\ln \bigl( K S_K + \sqrt{1 + K^2 S_K^2} \bigr)}{K} \bigr) 
    \ln \bigl( K S_K + \sqrt{1 + K^2 S_K^2} \bigr)^4  \biggr) 
\end{align}

and
\begin{equation}
\mathcal{D} = 8 K \pi^2 \sqrt{1 + K^2 S^2} \ln \left( K S + \sqrt{1 + K^2 S^2} \right)^3 \left( \ln \left( 4 \pi \right) - \ln \left( \frac{\ln \left( K S + \sqrt{1 + K^2 S^2} \right)}{K} \right) \right)^2
\end{equation}

\begin{figure}[h]	
	\centering
	\begin{subfigure}{0.37\textwidth}
		\includegraphics[width=\linewidth]{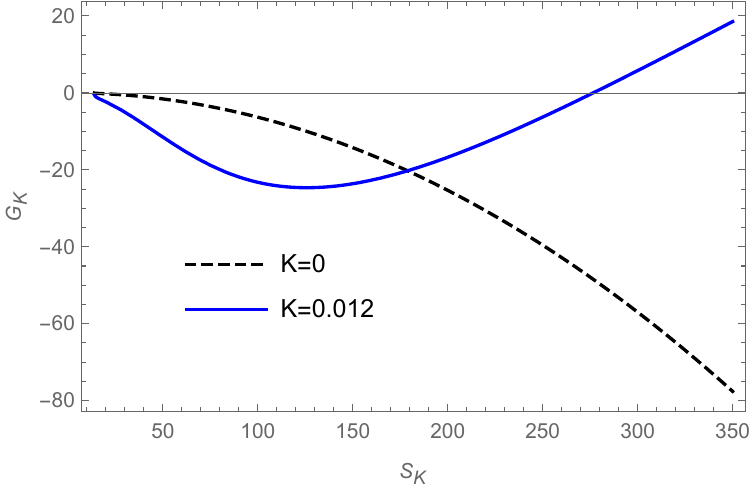}
		\caption{Gibbs free energy versus Kaniadakis entropy.}
		\label{15a}
		\end{subfigure}
		\begin{subfigure}{0.40\textwidth}
		\includegraphics[width=\linewidth]{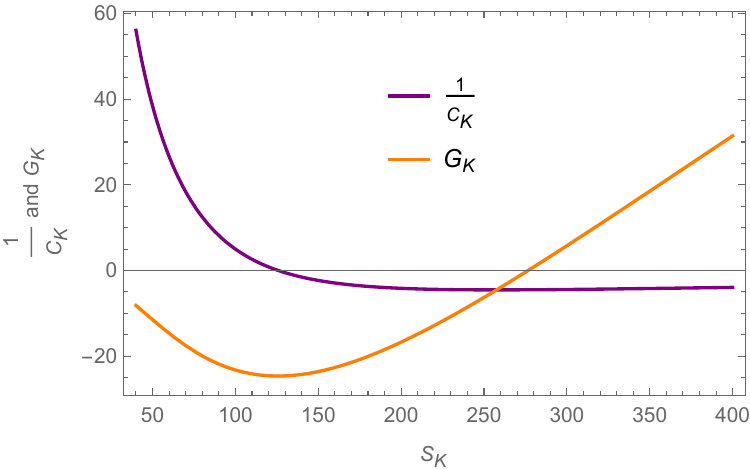}
		\caption{ Inverse specific heat and Gibbs free energy versus Kaniadakis entropy.}
		\label{15b}
		\end{subfigure}
		\caption{The Gibbs free energy curves for Kaniadakis entropy in  the fixed $(\Phi)$ ensemble for $l=1$ and $\Phi=0.2$.}
	\label{15}
 \end{figure}

In Fig.\ref{15a} the Gibbs free energy, $G_K$ is plotted against the Kaniadakis entropy, $S_K$ for $l=1$ and $\Phi=0.2$ in the fixed $(\Phi)$ ensemble. Here the solid blue curve represents the Gibbs free energy for the Kaniadakis parameter, $K=0.012$ whereas the black dashed line represents the free energy for the Bekenstein-Hawking(BH) entropy case $(K=0)$. The free energy for the $K=0$ case has been scaled down by multiplying it with $\frac{1}{10}$  whereas the free energy for the $K=0.012$ case is kept the same, this difference in scaling is just so that the distinct behaviour of both the curves could be seen more clearly. We keep in mind that the scaling does not affect the position or occurrence of the phase transition. We find that for $K=0.012$, the Gibbs free energy has a zero point at $S_K=276.64$ which indicates  the presence of a Hawking-Page phase transition present in the Kaniadakis modified black hole whereas the Gibbs free energy for the  BH entropy case shows no such behaviour. In  Fig.\ref{15b} we plot the inverse of specific heat capacity (purple) and the Gibbs free energy (orange) against the Kaniadakis entropy. The inverse of specific heat capacity  has been scaled up by multiplying it with $3000$  whereas the free energy is kept the same, this difference in scaling is just so that the distinct behaviour of both the curves could be seen more clearly. We keep in mind that the scaling does not affect the position or occurrence of the phase transitions. The points where these curves intersect the horizontal axis would represent the Davies-type (Purple) and Hawking-Page (Orange) phase transition respectively which are seen to be quite distant from each other.   \\	

		\subsection{The Renyi entropy case}
	
 \underline{\textbf{Fixed $(Q)$ ensemble}}: \\
  We replace the horizon radius in  (\ref{eqtn27}) by  the one obtained in (\ref{eqtn15}) to get the modified mass $M_{R}$ for the charged BTZ black hole which is given by :
  \begin{equation}
M_R = \frac{1 - 2 e^{\lambda S_R} + e^{2 \lambda S_R} + 8 \pi^3 Q^2 \lambda^2 \ln(4 \pi) - 8 \pi^3 Q^2 \lambda^2 ( \ln ( -1 + e^{\lambda S_R}) - \ln(\lambda))}{16 \pi^2 \lambda^2}
 \end{equation} \\

 The heat capacity for the Renyi entropy case in the fixed $(Q)$ ensemble can be calculated as:
$$C_{R}=  \frac{\frac{\partial M_{R}}{\partial S_{R}}}{\frac{\partial^2 M_{R}}{\partial S_{R} ^2}}\\= 
\frac{\left(-1 + e^{\lambda S_R}\right) \left(1 - 2 e^{\lambda S_R} + e^{2 \lambda S_R} - 
4 \pi^3 Q^2 \lambda^2\right)}{\lambda \left(-1 + 4 e^{\lambda S_R} - 5 e^{2 \lambda S_R} + 2 e^{3 \lambda S_R} + 
4 \pi^3 Q^2 \lambda^2\right)}$$\\

\begin{figure}[h]	
	\centering
	\begin{subfigure}{0.37\textwidth}
		\includegraphics[width=\linewidth]{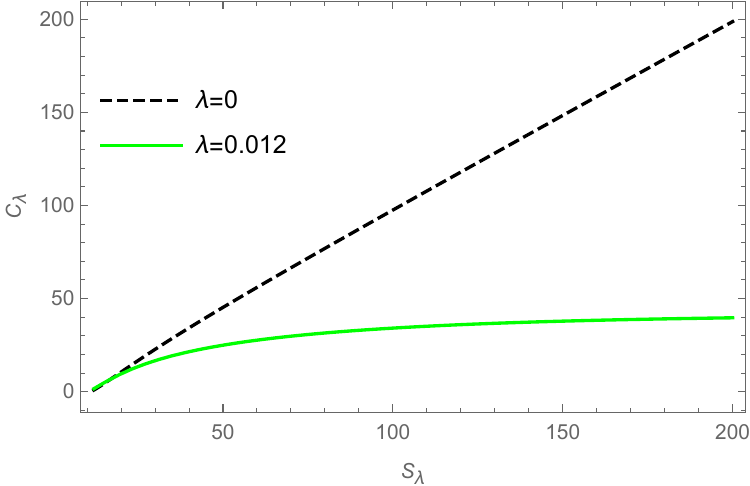}
		\caption{Heat capacity in the fixed $(Q)$ ensemble for $l=1$ and $Q=1$ }
		\label{16a}
		\end{subfigure}
		\begin{subfigure}{0.37\textwidth}
		\includegraphics[width=\linewidth]{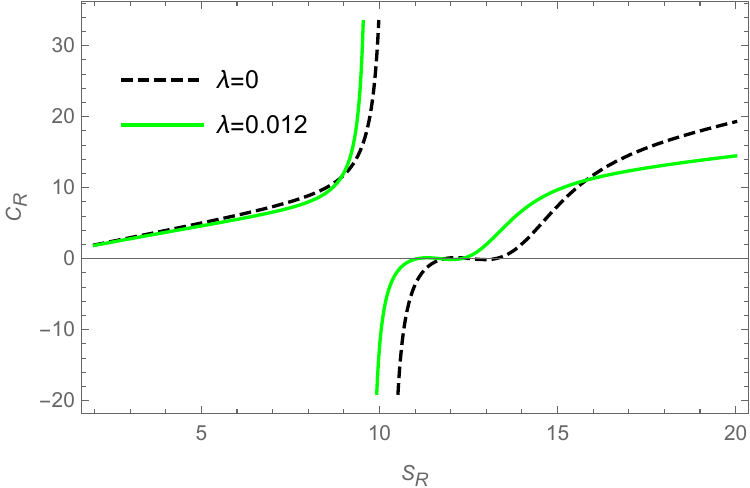}
		\caption{Heat capacity in the fixed $(\Phi)$ ensemble for $l=1$ and $\Phi=0.2$ }
		\label{16b}
		\end{subfigure}
		\caption{ The heat capacity versus Renyi entropy plots in the fixed $(J)$ and fixed $(\Phi)$ ensembles.}
	\label{16}
 \end{figure}

 We see from Fig.\ref{16a} that the heat capacity, $C_R$ for the  charged BTZ black hole is plotted against the Renyi entropy, $S_R$ in the fixed $(Q)$ ensemble. Here the solid green curve represents the heat capacity for the Renyi parameter, $\lambda=0.012$ whereas the black dashed line represents the heat capacity for the Bekenstein-Hawking(BH) entropy case $(\lambda=0)$. For the parameter $l=1$ and $ Q=1$ we find that there are no divergences in the heat capacity curve for both the Renyi and Bekenstein-Hawking (BH) entropy cases. \\
 
 The  Gibbs free energy for the Renyi entropy case in the fixed $(Q)$ ensemble can be obtained as:

$$G_{R} =   M_{R} - T_{R} S_{R}$$

\begin{align}
G_R = &\ \frac{1}{16 \pi^2 \lambda^2 (-1 + e^{\lambda S_R})} \Big(
    -1 + e^{3 \lambda S_R} (1 - 2 \lambda S_R) + e^{2 \lambda S_R} (-3 + 4 \lambda S_R)  - 8 \pi^3 Q^2 \lambda^2 \ln(4 \pi) 
    + e^{\lambda S_R} \big(3 \notag \\
&\ - 2 \lambda S_R + 8 \pi^3 Q^2 S_R \lambda^3 
    + 8 \pi^3 Q^2 \lambda^2 \ln(4 \pi) \big)  - 8 (-1 + e^{\lambda S_R}) \pi^3 Q^2 \lambda^2 
    \big( \ln ( -1 + e^{\lambda S_R}) - \ln(\lambda)\big) 
\Big).
\end{align}\\

\begin{figure}[h]	
	\centering
	\begin{subfigure}{0.37\textwidth}
		\includegraphics[width=\linewidth]{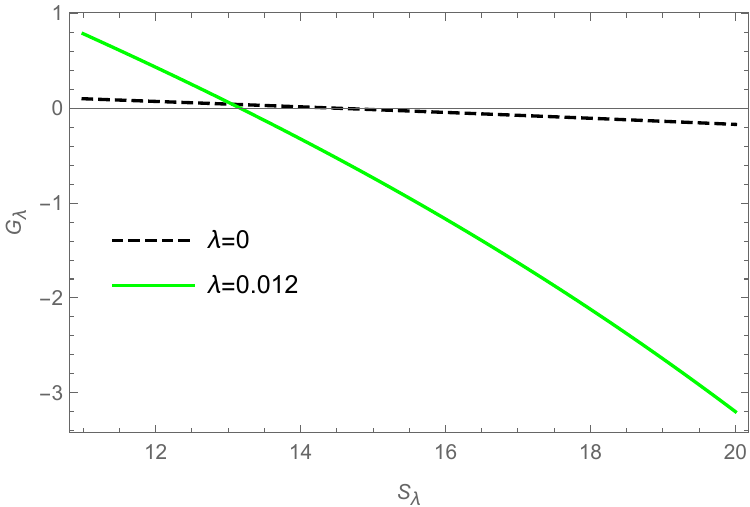}
		\caption{Free energy in the fixed $(Q)$ ensemble for $l=1$ and $Q=1$ }
		\label{17a}
		\end{subfigure}
		\begin{subfigure}{0.41\textwidth}
		\includegraphics[width=\linewidth]{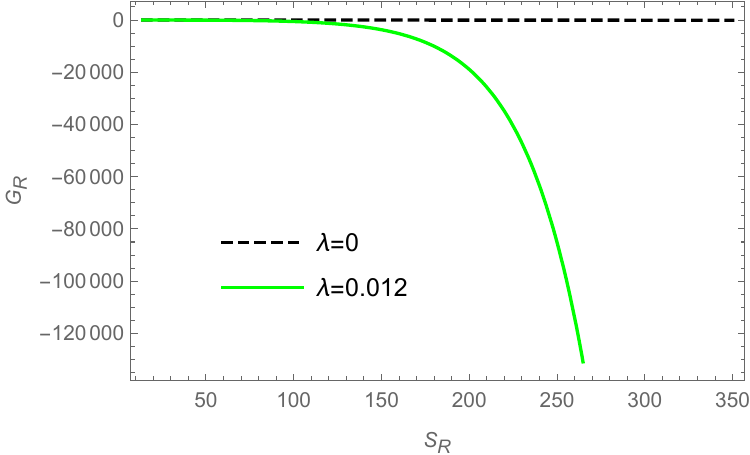}
		\caption{Free energy in  the fixed $(\Phi)$ ensemble for $l=1$ and $\Phi=0.2$.}
		\label{17b}
		\end{subfigure}
		\caption{ The Gibbs free energy versus Renyi entropy plots in the fixed $(Q)$ and fixed $(\Phi)$ ensembles.}
	\label{17}
 \end{figure}

In Fig.\ref{17a} the Gibbs free energy, $G_R$ is plotted against the Renyi entropy, $S_R$ for $l=1$ and $Q=1$ in the fixed $(Q)$ ensemble. Here the solid green curve represents the Gibbs free energy for the Renyi parameter, $\lambda=0.012$ whereas the black dashed line represents the free energy for the Bekenstein-Hawking(BH) entropy case $(\lambda=0)$. The free energy for the $\lambda=0$ case has been scaled down by multiplying it with $\frac{1}{10}$  whereas the free energy for the $\lambda=0.012$ case is kept the same, this difference in scaling is just so that the distinct behaviour of both the curves could be seen more clearly. We find that in Fig.\ref{19a} the Gibbs free energy goes to zero for both the entropy cases, where for $\lambda=0$, the zero point is at $S_R =14.55$ and for $\lambda=0.012$, the zero point is found to be at $S_R =13.17$, these zero points refer to the presence of a Hawking-Page phase transition present in the charged BTZ black hole for both the Bekenstein-Hawking as well as the Renyi entropy cases.\\	

 \underline{\textbf{Fixed $(\Phi)$ ensemble}}:\\
 
   The modified mass for the Renyi entropy case in the fixed $(\Phi)$ ensemble is given by:
 $$M_{R} ' = M_{R} - Q \frac{\partial M_{R}}{\partial Q}$$
 $$M_{R} ' = M_{R} - Q \Phi$$
 where $\frac{\partial M_{R}}{\partial Q}= \Phi$ is the electric potential of the black hole with respect to the electric charge $Q$. By making the necessary replacements and putting $l=1$, we get the final equation as follows:
\begin{equation}
M_R = \frac{-8 \pi \lambda^2 \Phi^2 + (-1 + e^{S_R \lambda})^2 \ln(4 \pi) 
      - (-1 + e^{S_R \lambda})^2 \big( \ln ( -1 + e^{\lambda S_R}) - \ln(\lambda)\big)}
      {16 \pi^2 \lambda^2 \left( \ln(4 \pi \lambda) - \big( \ln ( -1 + e^{\lambda S_R}\big) \right)}
\end{equation}\\

 We have here used the term $M_{R}$ instead of $M_{R}'$ for convenience. Here $M_{R}$ is the modified mass for the fixed $(\Phi)$ ensemble. The heat capacity for the Renyi entropy case in the fixed $(\Phi)$ ensemble can be calculated as:
$$C_{R}=  \frac{\frac{\partial M_{R}}{\partial S_{R}}}{\frac{\partial^2 M_{R}}{\partial S_{R} ^2}}\\=  \frac{A}{B}$$
where,
\begin{align}
A = &\ \bigl( -1 + e^{S_R \lambda} \bigr) 
      \biggl( 
        \ln(4 \pi) 
        - \ln \biggl( \frac{-1 + e^{S_R \lambda}}{\lambda} \biggr) 
      \biggr) \biggl( 
        -4 \pi \lambda^2 \Phi^2 
        + \bigl( -1 + e^{S_R \lambda} \bigr)^2 \ln(4 \pi)^2 \notag \\
    & \quad - 2 \bigl( -1 + e^{S_R \lambda} \bigr)^2 
              \ln(4 \pi) 
              \ln \biggl( \frac{-1 + e^{S_R \lambda}}{\lambda} \biggr) \quad + \bigl( -1 + e^{S_R \lambda} \bigr)^2 
              \ln \biggl( \frac{-1 + e^{S_R \lambda}}{\lambda} \biggr)^2  \biggr)
\end{align}\\

\begin{align}
B = &\ \lambda \biggl( 
        4 \pi \lambda^2 \Phi^2 \ln(4 \pi) 
        - \ln(4 \pi)^3 
        - 5 e^{2 S_R \lambda} \ln(4 \pi)^3  \quad + 2 e^{3 S_R \lambda} \ln(4 \pi)^3 
        + e^{S_R \lambda} \bigl( 
            -8 \pi \lambda^2 \Phi^2 
            + 4 \ln(4 \pi)^3 
          \bigr)\notag \\
    &   + \biggl( 
            -4 \pi \lambda^2 \Phi^2 
            - 3 \bigl( -1 + e^{S_R \lambda} \bigr)^2 \bigl( -1 + 2 e^{S_R \lambda} \bigr) 
              \ln(4 \pi)^2 
          \biggr) 
          \ln \biggl( \frac{-1 + e^{S_R \lambda}}{\lambda} \biggr)  + 3 \bigl( -1 + e^{S_R \lambda} \bigr)^2\notag \\
    & \bigl( -1 + 2 e^{S_R \lambda} \bigr) 
              \ln(4 \pi) 
              \ln \biggl( \frac{-1 + e^{S_R \lambda}}{\lambda} \biggr)^2  - \bigl( -1 + e^{S_R \lambda} \bigr)^2 
              \bigl( -1 + 2 e^{S_R \lambda} \bigr) 
              \ln \biggl( \frac{-1 + e^{S_R \lambda}}{\lambda} \biggr)^3 \biggr)
\end{align}

We see from Fig.\ref{16b} that the heat capacity, $C_R$ for the charged BTZ black hole is plotted against the Renyi entropy, $S_R$ in the fixed $(\Phi)$ ensemble. Here the solid green curve represents the heat capacity for the Renyi parameter, $\lambda=0.012$ whereas the black dashed line represents the heat capacity for the Bekenstein-Hawking(BH) entropy case $(\lambda=0)$. We see that for $l=1$ and $\Phi=0.2$ there is a divergence in the heat capacity curve for both the Renyi and BH entropy cases. The divergence for the Renyi entropy case $(K=0.012)$ is seen to be at $S_{K}=9.71$ whereas for the Bekenstein-Hawking entropy case $(K=0)$ the divergence is found to be at $S_{K}=10.24$.\\
    
    The  Gibbs free energy for the Renyi entropy case in the fixed $(\Phi)$ ensemble can be obtained as:

$$G_{R} =   M_{R} - T_{R} S_{R}$$

$$G_{R} = \frac{C}{D}$$
where,
\begin{align}
C = &\ \bigl( 
        8 \pi \lambda^2 \Phi^2 - \ln(4 \pi) 
      \bigr) \ln(4 \pi) 
      - e^{3 S_R \lambda} \bigl( -1 + 2 S_R \lambda \bigr) \ln(4 \pi)^2  + e^{2 S_R \lambda} \bigl( -3 + 4 S_R \lambda \bigr) \ln(4 \pi)^2 + e^{S_R \lambda} \biggl( 
          8 \pi \lambda^2 \Phi^2\notag \\
    &\ \bigl( S_R \lambda - \ln(4 \pi) \bigr)  + \bigl( 3 - 2 S_R \lambda \bigr) \ln(4 \pi)^2 
      \biggr)  + 2 \bigl( -1 + e^{S_R \lambda} \bigr) 
          \biggl( 
              4 \pi \lambda^2 \Phi^2 
              + \bigl( -1 + e^{S_R \lambda} \bigr) \bigl( 1 + e^{S_R \lambda}\notag \\
    &\ \bigl( -1 + 2 S_R \lambda \bigr) \bigr) 
                \ln(4 \pi) 
          \biggr) 
          \ln \biggl( \frac{-1 + e^{S_R \lambda}}{\lambda} \biggr)  - \bigl( -1 + e^{S_R \lambda} \bigr)^2 
          \bigl( 1 + e^{S_R \lambda} \bigl( -1 + 2 S_R \lambda \bigr) \bigr) 
          \ln \biggl( \frac{-1 + e^{S_R \lambda}}{\lambda} \biggr)^2
\end{align}\\

\begin{equation}
D = 16 (-1 + e^{S_R \lambda}) \pi^2 \lambda^2 \left( \ln(4 \pi) - \ln \left( \frac{-1 + e^{S_R \lambda}}{\lambda} \right) \right)^2
\end{equation}\\

In Fig.\ref{17b} the Gibbs free energy, $G_R$ is plotted against the Renyi entropy, $S_R$ for $l=1$ and $\Phi=0.2$ in the fixed $(\Phi)$ ensemble. Here the solid green curve represents the Gibbs free energy for the Renyi parameter, $\lambda=0.012$ whereas the black dashed line represents the free energy for the Bekenstein-Hawking(BH) entropy case $(\lambda=0)$.  The free energy for the $\lambda=0$ case has been scaled down by multiplying it with $\frac{1}{10}$  whereas the free energy for the $\lambda=0.012$ case is kept the same, this difference in scaling is just so that the distinct behaviour of both the curves could be seen more clearly. We find that unlike the Kaniadakis entropy there is no Hawking-Page phase transition present in the Renyi entropy case for the fixed $(\Phi)$ ensemble.\\

\subsection{The Barrow entropy case}
	
 \underline{\textbf{Fixed $(Q)$ ensemble}}: \\
  We replace the horizon radius in  (\ref{eqtn27}) by  the one obtained in (\ref{eqtn24}) to get the modified mass $M_{\Delta}$ for the charged BTZ black hole which is given by :
$$M_\Delta = \frac{{S_\Delta^{\frac{4}{2 + \Delta}} + 8 \pi^3 Q^2 \ln(4 \pi) - 8 \pi^3 Q^2 \ln\left(S_\Delta^{\frac{2}{2 + \Delta}}\right)}}{16 \pi^2}$$\\

 The heat capacity for the Barrow entropy case in the fixed $(Q)$ ensemble can be calculated as:
 $$C_{\Delta}=  \frac{\frac{\partial M_{\Delta}}{\partial S_{\Delta}}}{\frac{\partial^2 M_{\Delta}}{\partial S_{\Delta} ^2}}\\ =  \frac{\left(S_\Delta \left(4 \pi^3 Q^2 - S_\Delta^{\frac{4}{2 + \Delta}}\right) (2 + \Delta)\right)}{S_\Delta^{\frac{4}{2 + \Delta}} (-2 + \Delta) - 4 \pi^3 Q^2 (2 + \Delta)}
$$\\

\begin{figure}[h]	
	\centering
	\begin{subfigure}{0.37\textwidth}
		\includegraphics[width=\linewidth]{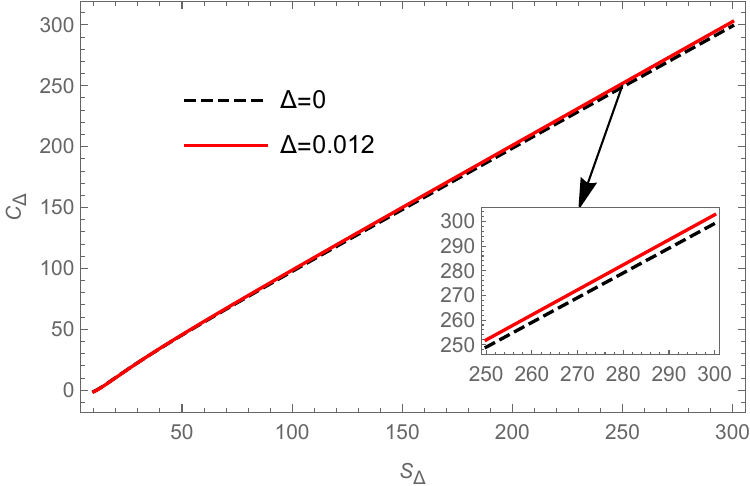}
		\caption{Heat capacity in the fixed $(Q)$ ensemble for $l=1$ and $Q=1$ }
		\label{18a}
		\end{subfigure}
		\begin{subfigure}{0.37\textwidth}
		\includegraphics[width=\linewidth]{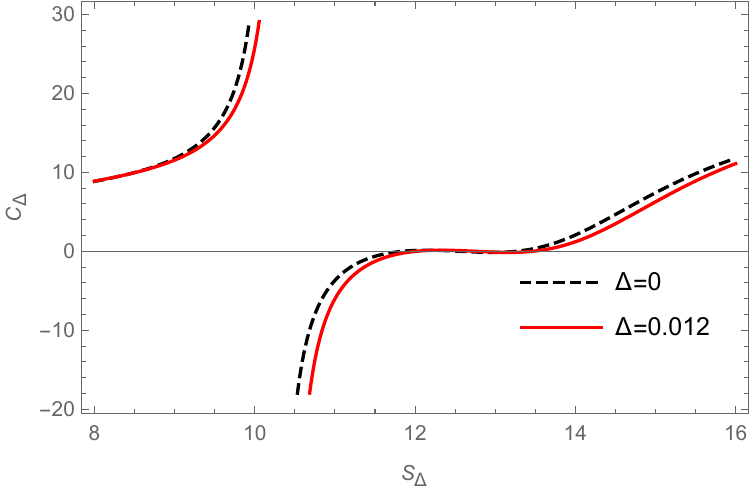}
		\caption{Heat capacity in the fixed $(\Phi)$ ensemble for $l=1$ and $\Phi=0.2$ }
		\label{18b}
		\end{subfigure}
		\caption{  The heat capacity versus Barrow entropy plots in the fixed $(J)$ and fixed $(\Phi)$ ensembles.}
	\label{18}
 \end{figure}

 We see from Fig.\ref{18a} that the heat capacity, $C_{\Delta}$ for the charged BTZ black hole is plotted against the Barrow entropy, $S_{\Delta}$ in the fixed $(Q)$ ensemble. Here the solid red curve represents the heat capacity for the Barrow parameter, $\Delta=0.012$ whereas the black dashed line represents the heat capacity for the Bekenstein-Hawking(BH) entropy case $(\Delta=0)$. For the parameter $l=1$ and $ Q=1$ we find that there are no divergences in the heat capacity curve for both the Barrow and Bekenstein-Hawking (BH) entropy cases.\\
 
  The  Gibbs free energy for the Renyi entropy case in the fixed $(Q)$ ensemble can be obtained as:

$$G_{\Delta} =   M_{\Delta} - T_{\Delta} S_{\Delta}$$

\begin{equation}
G_\Delta = \frac{S_\Delta^{\frac{4}{2 + \Delta}} (-2 + \Delta) + 8 \pi^3 Q^2 \left(2 + 2 \ln(4 \pi) + \Delta \ln(4 \pi)\right) - 8 \pi^3 Q^2 (2 + \Delta) \ln\left(S_\Delta^{\frac{2}{2 + \Delta}}\right)}{16 \pi^2 (2 + \Delta)}
\end{equation}\\

\begin{figure}[h]	
	\centering
	\begin{subfigure}{0.37\textwidth}
		\includegraphics[width=\linewidth]{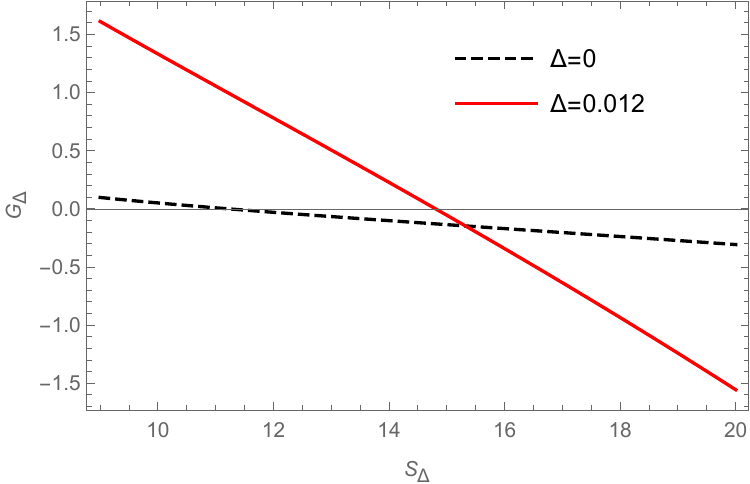}
		\caption{Free energy in the fixed $(Q)$ ensemble for $l=1$ and $Q=1$ }
		\label{19a}
		\end{subfigure}
		\begin{subfigure}{0.39\textwidth}
		\includegraphics[width=\linewidth]{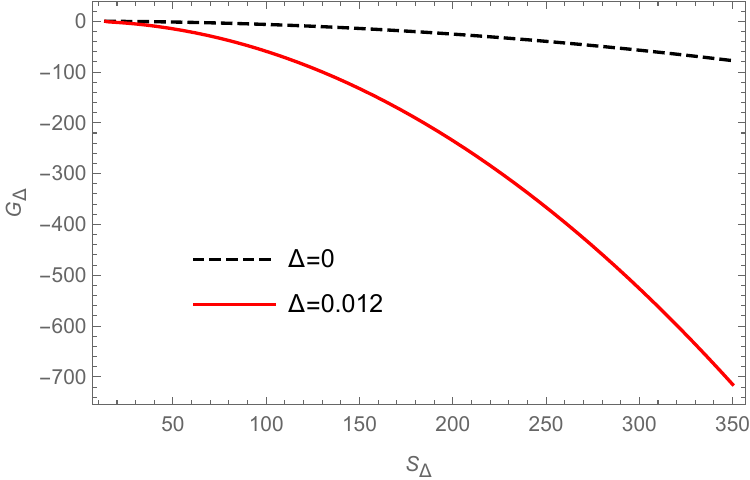}
		\caption{Free energy in  the fixed $(\Phi)$ ensemble for $l=1$ and $\Phi=0.2$.}
		\label{19b}
		\end{subfigure}
		\caption{ The Gibbs free energy versus Barrow entropy plots in the fixed $(Q)$ and fixed $(\Phi)$ ensembles.}
	\label{19}
 \end{figure}

In Fig.\ref{19a} the Gibbs free energy, $G_\Delta$ is plotted against the Barrow entropy, $S_\Delta$ for $l=1$ and $Q=1$ in the fixed $(Q)$ ensemble. Here the solid red curve represents the Gibbs free energy for the Barrow parameter, $\Delta=0.012$ whereas the black dashed line represents the free energy for the Bekenstein-Hawking(BH) entropy case $(\Delta=0)$.  The free energy for the $\Delta=0$ case has been scaled down by multiplying it with $\frac{1}{10}$  whereas the free energy for the $\Delta=0.012$ case is kept the same, this difference in scaling is just so that the distinct behaviour of both the curves could be seen more clearly. We find that in Fig.\ref{19a} the Gibbs free energy goes to zero for both the entropy cases, where for $\Delta=0$, the zero point is at $S_\Delta =14.55$ and for $\Delta=0.012$, the zero point is found to be at $S_\Delta =14.81$, these zero points refer to the presence of a Hawking-Page phase transition present in the charged BTZ black hole for both the Bekenstein-Hawking as well as the Barrow entropy cases.\\	
 
\underline{\textbf{Fixed $(\Phi)$ ensemble}}:\\

  The modified mass for the Barrow entropy case in the fixed $(\Phi)$ ensemble is given by:
 $$M_{\Delta} ' = M_{\Delta} - Q \frac{\partial M_{\Delta}}{\partial Q}$$
 $$M_{\Delta} ' = M_{\Delta} - Q \Phi$$
 where $\frac{\partial M_{\Delta}}{\partial Q}= \Phi$ is the electric potential of the black hole with respect to the electric charge $Q$. By making the necessary replacements and putting $l=1$, we get the final equation as follows:
 \begin{equation}	
M_\Delta = -\frac{8 \pi \Phi^2 - S_\Delta^{\frac{4}{2 + \Delta}} \ln(4 \pi) + S_\Delta^{\frac{4}{2 + \Delta}} \ln\left(S_\Delta^{\frac{2}{2 + \Delta}}\right)}{16 \pi^2 \left(\ln(4 \pi) - \ln\left(S_\Delta^{\frac{2}{2 + \Delta}}\right)\right)}
\end{equation}\\

 We have here used the term $M_{\Delta}$ instead of $M_{\Delta}'$ for convenience. Here $M_{\Delta}$ is the modified mass for the fixed $(\Phi)$ ensemble. The heat capacity for the Barrow entropy case in the fixed $(\Phi)$ ensemble can be calculated as:
 $$C_{\Delta}=  \frac{\frac{\partial M_{\Delta}}{\partial S_{\Delta}}}{\frac{\partial^2 M_{\Delta}}{\partial S_{\Delta} ^2}}\\= \frac{A}{B}$$
where,
\begin{equation}
A=  (2 S_\Delta + \Delta S_\Delta) \ln\left(\frac{\left(S_\Delta^{\frac{2}{2 + \Delta}}\right)}{4 \pi}\right) \left(-4 \pi \Phi^2 + S_\Delta^{\frac{4}{2 + \Delta}} \ln(4 \pi)^2 - 2 S_\Delta^{\frac{4}{2 + \Delta}} \ln(4 \pi) \ln\left(S_\Delta^{\frac{2}{2 + \Delta}}\right) + S_\Delta^{\frac{4}{2 + \Delta}} \ln\left(S_\Delta^{\frac{2}{2 + \Delta}}\right)^2\right)
\end{equation}
and
\begin{align}
B = &\ S_\Delta^{\frac{4}{2 + \Delta}} (-2 + \Delta) \ln(4 \pi)^3 
      - 4 \pi \Phi^2 \bigl( -4 + 2 \ln(4 \pi) + \Delta \ln(4 \pi) \bigr) + \biggl( 
          4 \pi (2 + \Delta) \Phi^2 
          - 3 S_\Delta^{\frac{4}{2 + \Delta}} (-2 + \Delta) \ln(4 \pi)^2 
      \biggr) \notag \\
    &\ \ln\bigl( S_\Delta^{\frac{2}{2 + \Delta}} \bigr)  + 3 S_\Delta^{\frac{4}{2 + \Delta}} (-2 + \Delta) \ln(4 \pi) 
      \ln\bigl( S_\Delta^{\frac{2}{2 + \Delta}} \bigr)^2 
     - S_\Delta^{\frac{4}{2 + \Delta}} (-2 + \Delta) 
      \ln\bigl( S_\Delta^{\frac{2}{2 + \Delta}} \bigr)^3
\end{align}

We see from Fig.\ref{18b} that the heat capacity, $C_{\Delta}$ for the charged BTZ black hole is plotted against the Barrow entropy, $S_{\Delta}$ in the fixed $(\Phi)$ ensemble. Here the solid red curve represents the heat capacity for the Barrow parameter, $\Delta=0.012$ whereas the black dashed line represents the free energy for the Bekenstein-Hawking(BH) entropy case $(\Delta=0)$.  We see that for $l=1$ and $\Phi=0.2$ there is a divergence in the heat capacity curve for both the Barrow and BH entropy cases. The divergence for the Barrow entropy case $(\Delta=0.012)$ is seen to be at $S_{\Delta}=10.37$ whereas for the Bekenstein-Hawking entropy case $(\Delta=0)$ the divergence is found to be at $S_{\Delta}=10.24$.\\

 The  Gibbs free energy for the Barrow entropy case in the fixed $(\Phi)$ ensemble can be obtained as:

$$G_{\Delta} =   M_{\Delta} - T_{\Delta} S_{\Delta}$$   

\begin{align}
G_\Delta = &\ \frac{1}{16 \pi^2 (2 + \Delta) 
        \bigl( \ln \bigl( \frac{S_\Delta^{\frac{2}{2 + \Delta}}}{4 \pi} \bigr) \bigr)^2} 
        \Bigg( 
          S_\Delta^{\frac{4}{2 + \Delta}} (-2 + \Delta) \ln(4 \pi)^2 - 8 \pi \Phi^2 \bigl( -2 + 2 \ln(4 \pi) + \Delta \ln(4 \pi) \bigr) \notag \\
    &\ \quad + 2 \biggl( 
          4 \pi (2 + \Delta) \Phi^2 
          - S_\Delta^{\frac{4}{2 + \Delta}} (-2 + \Delta) \ln(4 \pi) 
      \biggr) 
      \ln\bigl( S_\Delta^{\frac{2}{2 + \Delta}} \bigr)  + S_\Delta^{\frac{4}{2 + \Delta}} (-2 + \Delta) 
      \ln\bigl( S_\Delta^{\frac{2}{2 + \Delta}} \bigr)^2 
        \Bigg).
\end{align}\\

In Fig.\ref{19b} the Gibbs free energy, $G_\Delta$ is plotted against the Barrow entropy, $S_\Delta$ for $l=1$ and $\Phi=0.2$ in the fixed $(\Phi)$ ensemble. Here the solid red curve represents the Gibbs free energy for the Renyi parameter, $\Delta=0.012$ whereas the black dashed line represents the free energy for the Bekenstein-Hawking(BH) entropy case $(\Delta=0)$. Here the solid red curve represents the Gibbs free energy for the Barrow parameter, $\Delta=0.012$ whereas the black dashed line represents the free energy for the Bekenstein-Hawking(BH) entropy case $(\Delta=0)$. The free energy for the $\Delta=0$ case has been scaled down by multiplying it with $\frac{1}{10}$  whereas the free energy for the $\Delta=0.012$ case is kept the same, this difference in scaling is just so that the distinct behaviour of both the curves could be seen more clearly. We find that unlike the Kaniadakis entropy there is no Hawking-Page phase transition present in the Barrow entropy case for the fixed $(\Phi)$ ensemble.\\	

 
 \section{Thermodynamic geometry of the charged BTZ black hole}

We discuss the thermodynamic geometry  of the charged BTZ black hole for both the Ruppeiner and geometrothermodynamic (GTD) formalisms. The Ruppeiner metric is defined as the negative hessian of entropy with respect to other extensive variables. Therefore it is only the fixed $(Q)$ ensemble which holds good in the Ruppeiner formalism as  $Q$ is an extensive thermodynamic quantity whereas $\Phi$ is not.\\

 Therefore the fixed $(\Phi)$ ensemble do not hold good as per the definition of the Ruppeiner metric. We therefore discuss the thermodynamic geometry of  the fixed $(\Phi)$ ensemble in the GTD formalism which is defined on a multi-dimensional phase space that comprises of both the extensive and intensive  variables of the thermodynamic system. The GTD formalism is therefore the appropriate geometric formalism for discussing both the thermodynamic ensembles.\\
 
 We first discuss the Ruppeiner geometry of the charged BTZ black hole in the fixed $(Q)$ ensemble where we observe  a curvature singularity in the Ruppeiner scalar curve only for the Kaniadakis entropy case. We then discuss the GTD geometry of the black hole for both the thermodynamic ensembles, where we observe curvature singularities in the GTD scalar curves in Kaniadakis entropy case only. These curvature singularities are seen to occur exactly at the points where we saw divergences in the corresponding heat capacity curves in the previous section.
 
\subsection{\textbf{Ruppeiner formalism}}
 The Ruppeiner metric for the C-BTZ black hole in the fixed $(Q)$ ensemble  can be obtained from the relation (\ref{eqtn2}) and is given as :-
$$dS_{R}^{2}= \frac{1}{T} \left(\frac{\partial^2 M}{\partial S^2}dS^2 + \frac{\partial^2 M}{\partial Q^2} dQ^2  + 2\frac{\partial^2 M}{\partial S \partial Q} dS dQ\right) $$
 where  $M$, $S$, $T$ and $Q$ are the mass, entropy, Hawking Temperature and electric charge for the charged BTZ black hole respectively.\\
 
From the above metric we calculate the Ruppeiner scalar for all the non-extensive entropy cases in the fixed $(Q)$ ensemble. We find that there is a curvature singularity in the Ruppeiner scalar curve \textbf{only for the Kaniadakis entropy case}. For all the other entropy cases namely: the Renyi and Barrow entropy cases the Ruppeiner scalar curves remain regular with no singularities therein.  We have not presented here either the derivation or the final expressions for the Ruppeiner scalar due to their considerable length. However obtaining these expressions is fairly straight forward and involves only routine mathematical computations. We present a detailed analysis of the Ruppeiner thermodynamic geometry for the Kaniadakis entropy case in the  fixed $(Q)$ ensemble as follows:\\ 

\begin{figure}[h!t]
  \centering
		\includegraphics[width=9cm,height=6cm]{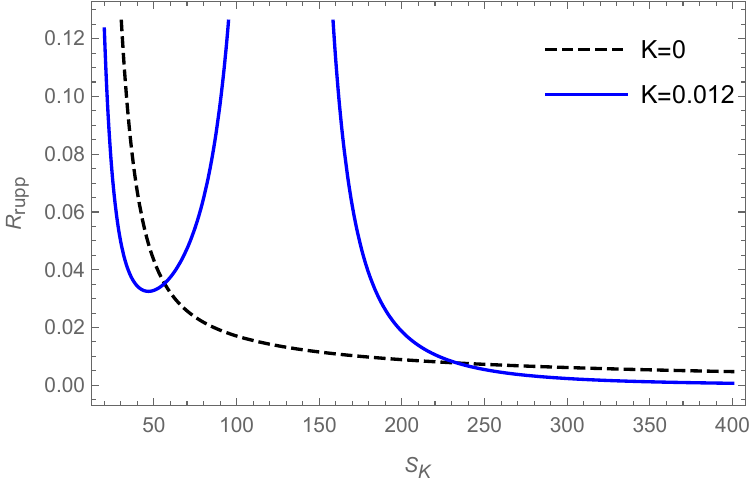}
		\caption{The Ruppeiner scalar versus Kaniadakis entropy plot in the fixed $(Q)$ ensemble for $l=1$ and $Q=1$.}
		\label{20}
		\end{figure}

We see from Fig.\ref{20} that the Ruppeiner scalar $R_{rupp}$ is plotted against the Kaniadakis entropy $S_K$ in the fixed $(Q)$ ensemble.  We see that for $l=1$ and $Q=1$, for the Kaniadakis parameter $K=0.012$ the Ruppeiner scalar curve has  a curvature singularity specifically at $S_K = 128.82$, which is depicted by the solid blue curve in the figure. The point of singularity here unlike the rotating BTZ black hole is similar to the point of divergence  obtained in the corresponding heat capacity curve for the Kaniadakis entropy case  in the fixed $(Q)$ ensemble. For the $K=0$ case however the Ruppeiner scalar curve is found to be regular everywhere with no curvature singularities as is depicted in the figure by the black dashed curve.     
 \\
		
\subsection{\textbf{Geometrothermodynamic(GTD) formalism}}
To describe the  charged BTZ black hole in GTD formalism, we first assume a five-dimensional phase space $\mathcal{T}$ with co-ordinates $M, S, Q, T$ and $\Phi$  which are the mass, entropy, electric charge,  temperature and electric potential  for the black hole. The contact 1-form for these set of co-ordinates can be written as:-
$$\Theta = dM -TdS -\Phi dQ $$
The 1-form satisfies the condition $\Theta \wedge (d\Theta)^3 \neq 0$ and a legendre invariant metric that goes by :-
$$G = (dM -TdS -\Phi dQ )^2 + TS(-dT dS + d\Phi dQ )$$
We then assume a two-dimensional sub-space of $\mathcal{T}$ with co-ordinates $E^a$ where $a=1,2$. We denote this subspace with $\epsilon$ which is defined as a smooth mapping, $\phi: \epsilon \rightarrow \mathcal{T}$. The subspace $\epsilon$ becomes the space of equilibrium states if the pullback of $\phi$ is 0 i.e. $\phi^{*}(\Theta)=0$. A metric structure $g$ is then naturally induced on $\epsilon$ by applying the same pullback on the metric G of $\mathcal{T}$. This induced metric determines all the geometric properties of the equilibrium space for the charged BTZ black hole.
\subsubsection{\textbf{The Kaniadakis entropy case}}
\underline{{\textbf{Fixed $(Q)$ ensemble}:}}\\

We write the GTD metric for the Kaniadakis entropy case in the fixed $(Q)$ ensemble from the general metric given in  (\ref{eqtn3}). We consider the thermodynamic potential $\varphi$ to be the mass $M_K$ for the Kaniadakis entropy case in  the fixed $(Q)$ ensemble as obtained in equation (\ref{eqtn11}). The GTD metric is given by :- 
$$g  =S_K \left(\frac{\partial M_K}{\partial S_K}\right)\left(- \frac{\partial^2 M_K}{\partial S_K^2} dS_K^2 + \frac{\partial^2 M_K}{\partial Q^2} dQ^2 \right) $$  
\\

From the above metric we calculate the GTD scalar for  the Kaniadakis entropy case in the fixed $(Q)$ ensemble. We have not presented here either the derivation or the final expressions for the GTD scalar due to their considerable length. However obtaining these expressions is fairly straight forward and involves only routine mathematical computations. We present a detailed analysis of the GTD thermodynamic geometry for the Kaniadakis entropy case in the  fixed $(Q)$ ensemble as follows:\\

\begin{figure}[h!t]
  \centering
		\includegraphics[width=9cm,height=6cm]{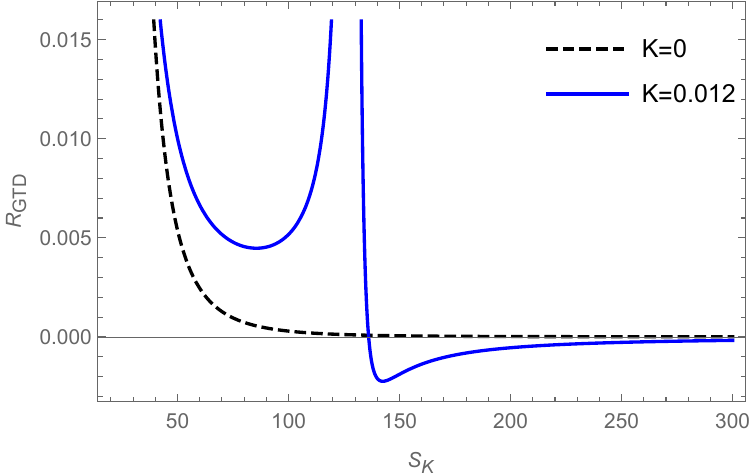}
		\caption{The GTD scalar versus Kaniadakis entropy plot in the fixed $(Q)$ ensemble for $l=1$ and $Q=1$.}
		\label{21}
		\end{figure}

We see from Fig.\ref{21} that the GTD scalar $R_{GTD}$ is plotted against the Kaniadakis entropy $S_K$  for the Kaniadakis parameter $K=0.012$ in the fixed $(Q)$ ensemble. We find that for  $l=1$ and $Q=1$, the GTD scalar has a curvature singularity at $S_K = 128.82$ for $K=0.012$ as depicted by the blue solid curve whereas for the BH entropy case ($K=0$) the curve remains regular  everywhere with no curvature singularities, as is depicted in the figure by the black dashed curve. The singularity obtained from the GTD scalar curve matches exactly with the point of divergence obtained from the corresponding heat capacity curve for the Kaniadakis entropy case in the fixed $(Q)$ ensemble.
 \\
		
		\underline{{\textbf{Fixed $(\Phi)$ ensemble}:}}\\

We write the GTD metric for the Kaniadakis entropy case in the fixed $(\Phi)$ ensemble from the general metric given in  (\ref{eqtn3}). We consider the thermodynamic potential $\varphi$ to be the mass $M_K$ for the Kaniadakis entropy case in  the fixed $(\Phi)$ ensemble as obtained in equation (\ref{eqtn14}). The GTD metric is given by :- 
$$g  =S_K \left(\frac{\partial M_K}{\partial S_K}\right)\left(- \frac{\partial^2 M_K}{\partial S_K^2} dS_K^2 + \frac{\partial^2 M_K}{\partial \Phi^2} d\Phi^2 \right) $$  
\\

From the above metric we calculate the GTD scalar for  the Kaniadakis entropy case in the fixed $(\Phi)$ ensemble. We have not shown here either the derivation or the final expressions for the GTD scalar due to their considerable length. However obtaining them is fairly straight forward and involves only routine mathematical computations. We present a detailed analysis of the GTD thermodynamic geometry for the Kaniadakis entropy case in the  fixed $(\Phi)$ ensemble as follows:\\

\begin{figure}[h]	
	\centering
	\begin{subfigure}{0.37\textwidth}
		\includegraphics[width=\linewidth]{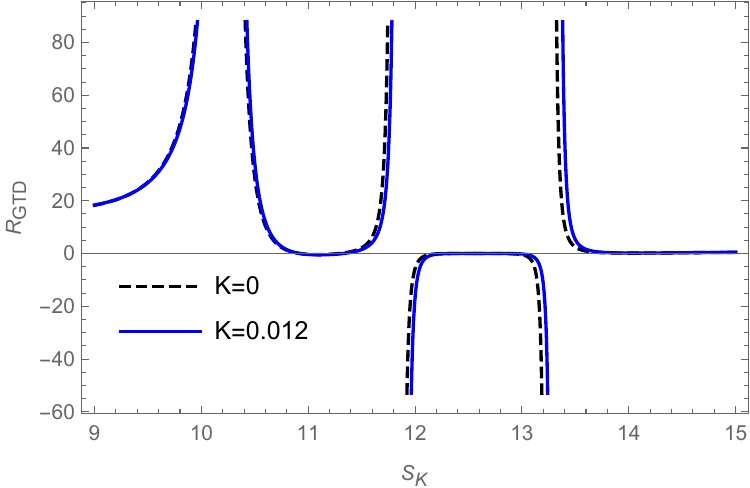}
		\caption{GTD scalar versus Kaniadakis entropy $(S \leq 15)$.}
		\label{22a}
		\end{subfigure}
		\begin{subfigure}{0.40\textwidth}
		\includegraphics[width=\linewidth]{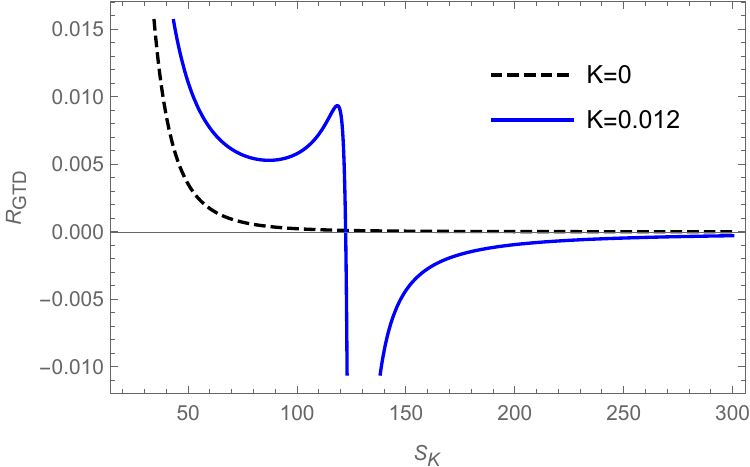}
		\caption{ GTD scalar versus Kaniadakis entropy $(15< S \leq 300)$.}
		\label{22b}
		\end{subfigure}
		\caption{The GTD scalar versus Kaniadakis entropy plot in  the fixed $(\Omega)$ ensemble  for $l=1$ and $\Phi=0.2$}
	\label{22}
 \end{figure}

In Fig.\ref{22} the GTD scalar, $R_{GTD}$ for the charged  BTZ black hole is plotted against the Kaniadakis entropy, $S_K$ in the fixed $(\Phi)$ ensemble. We draw the same plot in two different entropy ranges so as to make all the appearing singularities in the GTD scalar curve visible. We see from  Fig.\ref{22a} that for $l=1$ and $\Phi=0.2$ there is a singularity at $S_K=10.26$ for $K=0.012$ as shown by the solid blue curve. The heat capacity in the Bekenstein-Hawking(BH) entropy case ($K=0)$ here also produces a singularity at $S_K=10.24$  as shown by the black dashed curve. Although we can see the presence of two additional curvature singularities observed in the figure for both the Kaniadakis and BH entropy cases, however it was found that these additional curvature singularities correspond to the points where the heat capacity goes to zero for both the Kaniadakis and BH entropy cases in the fixed $(\Phi)$ ensemble. In Fig.\ref{22b} we see that there is a singularity in the heat capacity curve at $S_K=125.74$ for $K=0.012$ whereas here for $(K=0)$ case no such behaviour is observed in the GTD scalar curve.
\\ 	

		\subsubsection{\textbf{The Renyi entropy case}}
\underline{{\textbf{Fixed $(\Phi)$ ensemble}:}}\\

We write the GTD metric for the Renyi entropy case in the fixed $(\Phi)$ ensemble from the general metric given in  (\ref{eqtn3}). We consider the thermodynamic potential $\varphi$ to be the mass $M_K$ for the Renyi entropy case in  the fixed $(\Phi)$ ensemble. The GTD metric is given by :- 
$$g  =S_R \left(\frac{\partial M_R}{\partial S_R}\right)\left(- \frac{\partial^2 M_R}{\partial S_R^2} dS_R^2 + \frac{\partial^2 M_R}{\partial \Phi^2} d\Phi^2 \right) $$  
\\

From the above metric we calculate the GTD scalar for  the Renyi entropy case in the fixed $(\Phi)$ ensemble. We have not shown here either the derivation or the final expressions for the GTD scalar due to their considerable length. However obtaining them is fairly straight forward and involves only routine mathematical computations. We present a detailed analysis of the GTD thermodynamic geometry for the Renyi entropy case in the  fixed $(\Phi)$ ensemble as follows:\\

\begin{figure}[h!t]
  \centering
		\includegraphics[width=9cm,height=6cm]{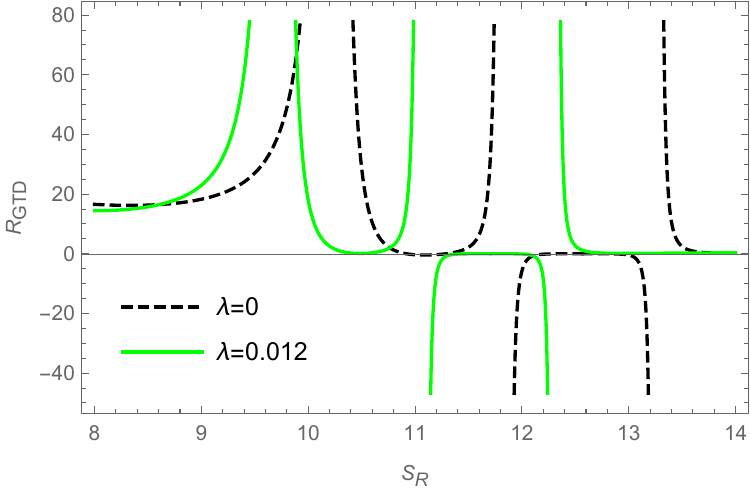}
		\caption{The GTD scalar versus Renyi entropy plot in  the fixed $(\Phi)$ ensemble  for $l=1$ and $\Phi=0.2$}
		\label{23}
		\end{figure}

We see from Fig.\ref{23} that the GTD scalar, $R_{GTD}$ for the charged  BTZ black hole is plotted against the Renyi entropy, $S_R$ in the fixed $(\Phi)$ ensemble.  We see that for $l=1$ and $\Phi=0.2$ there are curvature singularities in the GTD scalar curve for both the Renyi $(\lambda=0.012)$ and Bekenstein-Hawking (BH) entropy $(\lambda=0)$ cases. We find that  for $\lambda=0.012$ there is a curvature singularity observed in the GTD scalar curve at $S_R=9.71$ as depicted by the solid green curve in the figure. For the BH entropy  case ($\lambda=0$) we find that the GTD scalar has a curvature singularity at $S_R=10.24$   as shown in the figure by the black dashed curve. These singularities match exactly with the points of Davies type phase transitions observed in the corresponding heat capacity curve in the fixed $(\Phi)$ ensemble. There are two additional curvature singularities observed in the figure for both the Renyi and BH entropy cases, however it was found that these additional curvature singularities correspond to the points where the heat capacity goes to zero for both the Renyi and BH entropy cases in the fixed $(\Phi)$ ensemble.
		
		\subsubsection{\textbf{The Barrow entropy case}}
\underline{{\textbf{Fixed $(\Phi)$ ensemble}:}}\\

We write the GTD metric for the Barrow entropy case in the fixed $(\Phi)$ ensemble from the general metric given in  (\ref{eqtn3}). We consider the thermodynamic potential $\varphi$ to be the mass $M_{\Delta}$ for the Barrow entropy case in  the fixed $(\Phi)$ ensemble. The GTD metric is given by :- 
$$g  =S_{\Delta} \left(\frac{\partial M_{\Delta}}{\partial S_{\Delta}}\right)\left(- \frac{\partial^2 M_{\Delta}}{\partial S_{\Delta}^2} dS_{\Delta}^2 + \frac{\partial^2 M_{\Delta}}{\partial \Phi^2} d\Phi^2 \right) $$  
\\

From the above metric we calculate the GTD scalar for  the Barrow entropy case in the fixed $(\Phi)$ ensemble. We have not shown here either the derivation or the final expressions for the GTD scalar due to their considerable length. However obtaining them is fairly straight forward and involves only routine mathematical computations. We present a detailed analysis of the GTD thermodynamic geometry for the Barrow entropy case in the  fixed $(\Phi)$ ensemble as follows:\\

\begin{figure}[h!t]
  \centering
		\includegraphics[width=9cm,height=6cm]{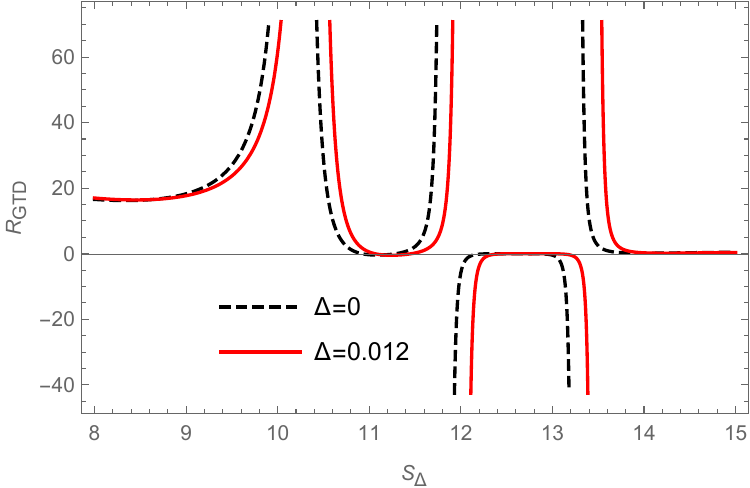}
		\caption{The GTD scalar versus Barrow entropy plot in  the fixed $(\Phi)$ ensemble  for $l=1$ and $\Phi=0.2$}
		\label{24}
		\end{figure}

We see from Fig.\ref{24} that the GTD scalar, $R_{GTD}$ for the charged  BTZ black hole is plotted against the Barrow entropy, $S_{\Delta}$ in the fixed $(\Phi)$ ensemble.  We see that for $l=1$ and $\Phi=0.2$ there are curvature singularities in the GTD scalar curve for both the Barrow $(\Delta=0.012)$ and Bekenstein-Hawking (BH) entropy $(\Delta=0)$ cases. We find that  for $\Delta=0.012$ there is a curvature singularity observed in the GTD scalar curve at $S_{\Delta}=10.37$ as depicted by the solid red curve in the figure. For the BH entropy  case ($\Delta=0$) we find that the GTD scalar has a curvature singularity at $S_{\Delta}=10.24$   as shown in the figure by the black dashed curve. These singularities match exactly with the points of Davies type phase transitions observed in the corresponding heat capacity curve in the fixed $(\Phi)$ ensemble. There are two additional curvature singularities observed in the figure for both the Barrow and BH entropy cases, however it was found that these additional curvature singularities correspond to the points where the heat capacity goes to zero for both the Barrow and BH entropy cases in the fixed $(\Phi)$ ensemble. 
\\

 \section{Conclusions:}  
We investigated the thermodynamics and thermodynamic geometry of the rotating(R-BTZ) and charged (C-BTZ) Banados-Teitelboim-Zanelli  black hole in three different non extensive entropy formalisms  namely: the Kaniadakis entropy, Renyi entropy and Barrow entropy respectively. We investigated all the thermodynamic ensembles of these black holes namely: the fixed $(J)$ and fixed $(\Omega)$ ensemble for the rotating BTZ black hole and the fixed $(Q)$ and fixed $(\Phi)$ ensemble for the charged BTZ black hole. We find that there are Davies type along with the Hawking-Page phase transitions in both the black holes observed in the Kaniadakis entropy case for all the thermodynamic ensembles. We found that both the Ruppeiner and the GTD scalar were observed to have curvature singularities corresponding to Davies type of phase transitions in the corresponding cases. The location of the curvature singularities in the GTD scalar exactly matched the Davies type phase transition points in both the ensembles. The Ruppeiner scalar, although showed curvature singularities, the location of those singularities were found to be not exactly the same as the corresponding Davies type transition points for the R-BTZ black hole. For all the other non-extensive entropy cases, the thermodynamic structure were found to resemble their counterparts in the Bekenstein-Hawking entropy. A table each for the two black holes are shown below summarising our results where we write the \textbf{number of  phase transition points} (both Davies type and Hawking-Page) observed against each of the non extensive entropy cases along with the BH entropy for all the thermodynamic ensembles. The table representing the R-BTZ black hole is as follows:
 \begin{table}[h!]
    \centering
    \begin{tabular}{|c|c|c|c|c|}
        \hline
        \textbf{Entropy} & \multicolumn{2}{c|}{\textbf{Fixed (J) ensemble}} & \multicolumn{2}{c|}{\textbf{Fixed (\(\Omega\)) ensemble}} \\
        \hline
        & \textbf{Davies-type} & \textbf{Hawking-Page} & \textbf{Davies-type} & \textbf{Hawking-Page} \\
        & \textbf{phase transition} & \textbf{phase transition} & \textbf{phase transition} & \textbf{phase transition} \\
        \hline
        Bekenstein-Hawking & 0 & 1 & 0 & 0 \\
        \hline
        Kaniadakis & 1 & 2 & 1 & 1 \\
        \hline
        Renyi & 0 & 1 & 0 & 0 \\
        \hline
        Barrow & 0 & 1 & 0 & 0 \\
        \hline
    \end{tabular}
\end{table}

The table representing the C-BTZ black hole is as follows:
 \begin{table}[h!]
    \centering
    \begin{tabular}{|c|c|c|c|c|}
        \hline
        \textbf{Entropy} & \multicolumn{2}{c|}{\textbf{Fixed (Q) ensemble}} & \multicolumn{2}{c|}{\textbf{Fixed (\(\Phi\)) ensemble}} \\
        \hline
        & \textbf{Davies-type} & \textbf{Hawking-Page} & \textbf{Davies-type} & \textbf{Hawking-Page} \\
        & \textbf{phase transition} & \textbf{phase transition} & \textbf{phase transition} & \textbf{phase transition} \\
        \hline
        Bekenstein-Hawking & 0 & 1 & 1 & 0 \\
        \hline
        Kaniadakis & 1 & 2 & 2 & 1 \\
        \hline
        Renyi & 0 & 1 & 1 & 0 \\
        \hline
        Barrow & 0 & 1 & 1 & 0 \\
        \hline
    \end{tabular}
\end{table}

	The key distinction of Kaniadakis entropy lies in its impact on the behaviour of black hole temperature. In contrast to the other considered entropic formalisms, where temperature is a monotonically increasing function of entropy, Kaniadakis entropy introduces a different thermal evolution. We have observed that in the Kaniadakis entropy framework, black hole temperature rises to a maximum value before gradually declining and asymptotically tending to zero for larger values of entropy. This non-monotonic behaviour in the temperature is what gives rise to the advent of the phase transitions (both Davies type and Hawking-Page) in the specific heat capacity and the Gibbs free energy respectively for the rotating and charged BTZ black holes. The behaviour of temperature in the R-BTZ black hole for both the Kaniadakis and Bekenstein Hawking entropy case is depicted in Fig.\ref{12}.\\

We know that the specific heat capacity of a black hole is given by, $C=\frac{T}{(\frac{\partial T}{\partial S})}$. Thus, when the temperature attains its maximum value, the slope of the graph i.e. $\frac{\partial T}{\partial S}$ equals zero and the specific heat capacity undergoes a divergence. When the temperature starts decreasing, the slope of the T-S graph becomes negative and so does the heat capacity. We also know that the black hole temperature is a first order derivative of mass with respect to entropy i.e. $T=(\frac{\partial M}{\partial S})$ where the mass M for the rotating BTZ black hole is given by:
$$M=  \frac{4 \pi^2 J^2}{S_{BH}^2} + \frac{S_{BH}^2}{16 \pi^2 l^2}$$
where after writing the Bekenstein-Hawking entropy, $S_{BH}$ in terms of  the Kaniadakis entropy i.e. $S_{BH} = \frac{Arc Sinh(K S)}{ K}$ gives the Kaniadakis corrected mass which is given by:

$$M= \frac{4 \pi^2 K^2 J^2}{ArcSinh^2 [K S]} + \frac{ArcSinh^2 [K S]}{16 \pi^2 l^2 K^2} $$

It is actually the second term in the above equation which is responsible for the unusual behaviour in the the temperature of the black hole. Here, the term $ \frac{ArcSinh^2 [K S]}{16 \pi^2 l^2 K^2}$ when differentiated with respect to the entropy in order to obtain the temperature becomes $ \frac{ArcSinh [K S]}{ 8 \pi l K \sqrt{1 + K^2 S^2}}$. This term is directly responsible for the dipping behaviour in the temperature because here as 
 $\frac{ArcSinh [K S]}{K}$ is an increasing function of entropy, the term $\frac{1}{\sqrt{1 + K^2 S^2}}$ decreases with an increase in entropy and thus they together produce this unusual behaviour in the temperature where it starts to fall after reaching a maximum value. The value of the maximum attainable temperature for a given set of $(J,l)$ values is fixed by the kaniadakis parameter,K as is evident from the above analysis. The non trivial temperature profile in the Kaniadakis entropy formalism is also responsible for the Hawking-Page phase transition obtained from the Gibbs free energy of the black hole. The Gibbs free energy is given by $G = M - T S$ where the Gibbs free energy is negative for small entropy values but as temperature starts falling after having attained its maximum value, $G$ starts to increase and finally reaches a zero point. On further increase in entropy the free energy becomes positive and thus the black hole becomes thermodynamically unstable as can be seen from Fig.\ref{2} and Fig.\ref{4}. The case is similar for the charged BTZ black hole where the Kaniadakis corrected mass for the C-BTZ black hole is given by:
 $$M_{K}= \frac{ArcSinh^2 [K S_{K}]}{16 \pi^2 l^2 K^2} - \frac{\pi Q^2}{2} \ln \biggl[\frac{ArcSinh[K S_{K}]}{4 \pi K}\biggr] $$
  It can be seen here as well that the term $ \frac{ArcSinh^2 [K S]}{16 \pi^2 l^2 K^2}$ when differentiated with respect to the entropy in order to obtain the temperature becomes $ \frac{ArcSinh [K S]}{ 8 \pi l K \sqrt{1 + K^2 S^2}}$. We therefore observe the dip in the temperature here as well which in turn results in the occurrence of a Davies type phase transition accompanied by a Hawking-Page phase transition in the charged BTZ black hole.\\
 
 This unusual behaviour of the temperature in the Kaniadakis entropy case seems to be inherent in the formulation of this entropy where it has been shown in \cite{kaniadakis2005statistical} that its mathematical structure is compatible with Einstein's special theory of relativity along with the second law of thermodynamics. It was also seen in \cite{kumar2024relativistic} where the author showed that the corrections in the black hole entropy obtained by Kaniadakis statistics in high temperature limit are indeed logarithmic in nature which prove that these corrections are indeed legitimate.   We could not find any specific reasons as to why a relativistic statistical approach to black holes should cause the temperature to behave so non-trivially. It would perhaps need a more sophisticated analysis of black holes in Kaniadakis formalism and also Kaniadakis entropy in general to find out why such non-triviality should occur.  \\
 \begin{figure}[h]	
	\centering
	\begin{subfigure}{0.37\textwidth}
		\includegraphics[width=\linewidth]{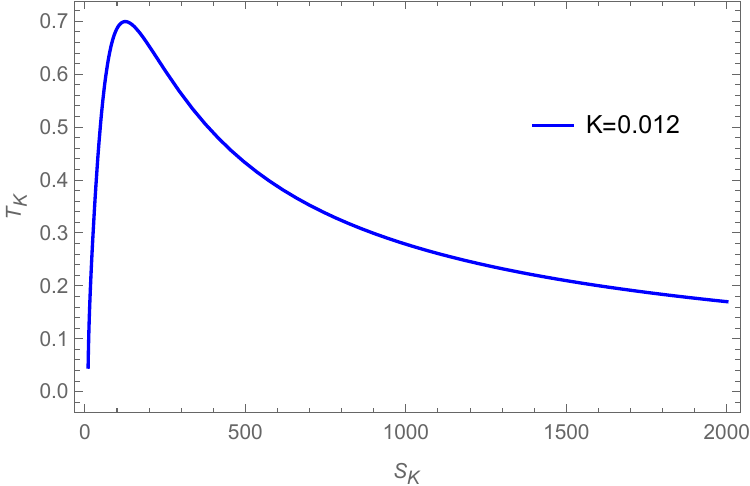}
		\caption{Temperature versus Kaniadakis entropy in the fixed $(J)$ ensemble .}
		\label{12a}
		\end{subfigure}
		\begin{subfigure}{0.37\textwidth}
		\includegraphics[width=\linewidth]{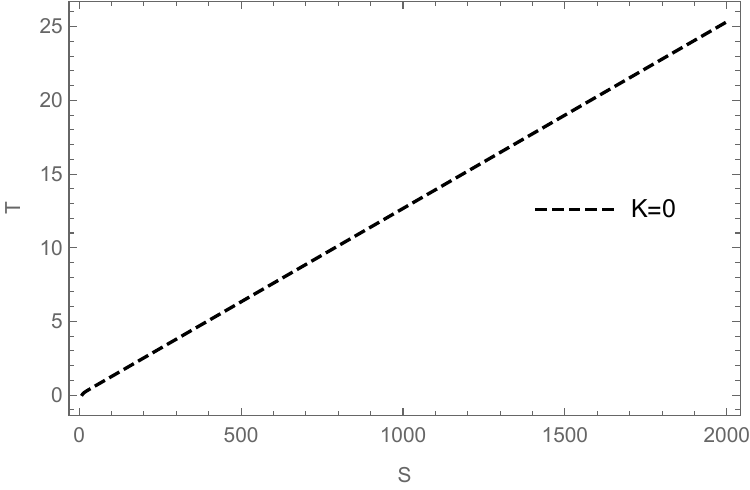}
		\caption{Temperature versus Bekenstein-Hawking entropy in the fixed $(J)$ ensemble.}
		\label{12b}
		\end{subfigure}
		\begin{subfigure}{0.37\textwidth}
		\includegraphics[width=\linewidth]{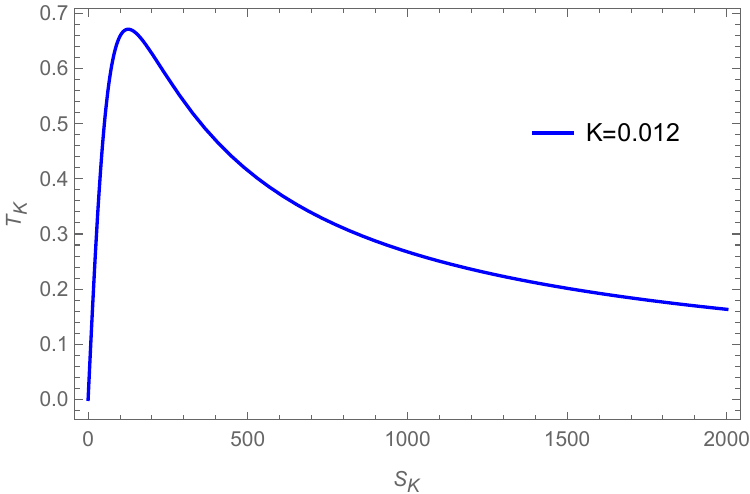}
		\caption{ Temperature versus Kaniadakis entropy in the fixed $(\Omega)$ ensemble.}
		\label{12c}
		\end{subfigure}
		\begin{subfigure}{0.37\textwidth}
		\includegraphics[width=\linewidth]{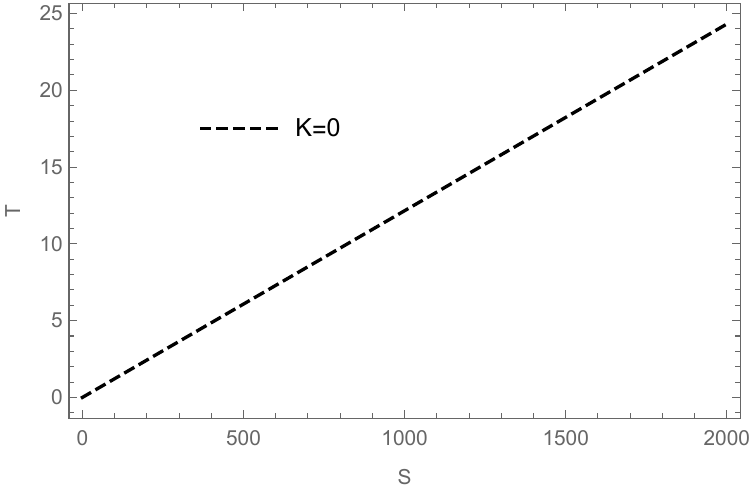}
		\caption{ Temperature versus Bekenstein-Hawking entropy in the fixed $(\Omega)$ ensemble.}
		\label{12d}
		\end{subfigure}
		\caption{The temperature curves for Kaniadakis entropy in  the fixed $(J)$ and $(\Omega)$ ensemble for $l=1,J=1$ and $\Omega=0.2$.}
	\label{12}
 \end{figure} 
 
	To conclude, our analysis shows that the thermodynamic phase structure on $(2+1)$ dimensional BTZ black hole with Kaniadakis entropy differs from the one with the conventional Bekenstein-Hawking entropy in all the ensembles. This result is in contrast to that found in \cite{luciano2023p}, in the context of charged AdS black holes where both Kaniadakis entropy and Bekenstein-Hawking entropy yielded similar phase structures. It will be interesting to extend the study of black hole thermodynamics in the framework of non-extensive entropy to other black holes to get a clearer picture. We plan to address this issue in our future works.

\section{Acknowledgments}
	The authors would like to thank Himanshu Bora for numerous enlightening discussions and certain exquisite suggestions he offered during the course of this work.
 
\bibliography{mybibfile2}

\end{document}